\documentclass[journal, compsoc]{article}

\usepackage[left=1.5cm, right=1.5cm, top=1.785cm, bottom=2.0cm]{geometry}

\usepackage[dvipsnames, table]{xcolor}
\usepackage{graphicx}
\usepackage{url}
\usepackage{tabularx}

\usepackage{amsmath,amsthm,amssymb}
\usepackage{mathtools}
\usepackage{booktabs}
\usepackage{enumitem}
\usepackage{bm}
\usepackage{comment}
\usepackage{dblfloatfix}
\usepackage{siunitx}
\usepackage{hyperref}
\usepackage{authblk}
\usepackage{makecell}

\usepackage{algorithm} 
\usepackage{algpseudocode}

\newcolumntype{s}{>{\columncolor[gray]{0.95}} p{1.7cm}}

\usepackage{etoolbox}
\usepackage{cite}

\newcommand{\revision}[1]{{\color{black} #1}}

\newcommand{\method}{MoINN}

\begin{document}

\title{\textbf{Automatic Identification of Chemical Moieties}}

\author[1,2]{Jonas~Lederer\thanks{Corresponding Authors: Jonas Lederer, Oliver T. Unke \\ \href{mailto:jonas.lederer@tu-berlin.com}{jonas.lederer@tu-berlin.com}, \href{mailto:oliver.unke@googlemail.com}{oliver.unke@googlemail.com}}}
\author[1,2]{Michael~Gastegger}
\author[1,2]{Kristof~T.~Schütt}
\author[3]{Michael~Kampffmeyer}
\author[1,2,4,5,6]{Klaus-Robert~Müller}
\author[1,2,4]{Oliver~T.~Unke$^*$}

\affil[1]{Berlin Institute of Technology (TU Berlin), 10587 Berlin, Germany}
\affil[2]{BIFOLD -- Berlin  Institute  for  the Foundations  of  Learning  and  Data, Germany}
\affil[3]{Department of Physics and Technology, UiT The Arctic University of Norway, 9019 Tromsø, Norway}
\affil[4]{Google Research, Brain team, Berlin.}
\affil[5]{Department of Artificial Intelligence, Korea University, Seoul 136-713, Korea}
\affil[6]{Max Planck Institut f{\"u}r Informatik, 66123 Saarbr{\"u}cken, Germany}

\date{\vspace{-7ex}}

\twocolumn[{%
  \begin{@twocolumnfalse}
    \maketitle
    \begin{abstract}
        In recent years, the prediction of quantum mechanical observables with machine learning methods has become increasingly popular. Message-passing neural networks (MPNNs) solve this task by constructing atomic representations, from which the properties of interest are predicted. Here, we introduce a method to automatically identify chemical moieties (molecular building blocks) from such representations, enabling a variety of applications beyond property prediction, which otherwise rely on expert knowledge. The required representation can either be provided by a pretrained MPNN, or learned from scratch using only structural information. Beyond the data-driven design of molecular fingerprints, the versatility of our approach is demonstrated by enabling the selection of representative entries in chemical databases, the automatic construction of coarse-grained force fields, as well as the identification of reaction coordinates.
        \vspace{0.5cm}
    \end{abstract}
  \end{@twocolumnfalse}
}]
{
  \renewcommand{\thefootnote}%
    {\fnsymbol{footnote}}
  \footnotetext[1]{Corresponding authors: Jonas Lederer, Oliver T. Unke\\ \href{mailto:jonas.lederer@tu-berlin.com}{jonas.lederer@tu-berlin.com}, \href{mailto:oliver.unke@googlemail.com}{oliver.unke@googlemail.com}}
}


\section{Introduction}

The computational study of structural and electronic properties of molecules is key to many discoveries in physics, chemistry, biology, and materials science. In this context, machine learning (ML) methods have become increasingly popular as a means to circumvent costly quantum mechanical calculations~\cite{behler_generalized_2007,bartok_gaussian_2010,rupp_fast_2012,schutt_schnet:_2017,chmiela_machine_2017,han_deep_2017,schutt_schnet_2018,zhang_deep_2018-1, schutt_schnetpack:_2018, sauceda2020construction,     unke2019physnet,schutt2021equivariant,unke2021machine,unke2021spookynet, klicpera2020directional, batzner20223, schutt_quantum-chemical_2017,noe2020machine,von2020exploring,keith2021combining, butler2018machine, chmiela2018towards, smith2017ani, popova2018deep, gebauer2019symmetry, gebauer2022inverse, chmiela2019sgdml, 10.1063/5.0138367,chmiela2023accurate, lederer2019machine, musaelian2023learning, gasteiger_dimenet_2020, gasteiger2020fast, doerr2021torchmd, huang2020quantum, huang2021ab, hansen2015machine}. One class of such ML methods are message passing neural networks (MPNNs)~\cite{gilmer_neural_2017}, which provide molecular property predictions based on end-to-end learned representations of atomic environments.

In contrast to such fine-grained representations, chemists typically characterize molecules by larger substructures (e.g.\ functional groups) to reason about their properties ~\cite{doi:10.1021/jm00120a002, duarte2007privileged, lemke2003review}. This gives rise to the idea of using MPNNs for the automatic identification of ``chemical moieties'', or characteristic parts of the molecule, to which its properties can be traced back. Since manually searching for moieties that explain (or are characteristic of) certain properties of molecules is a complex and tedious task, the capability of ML to find patterns and correlations in data could ease the identification of meaningful substructures drastically.

Previous work has introduced a variety of different approaches to identify substructures in molecules, with objectives ranging from substructure mining~\cite{ertl2017algorithm, klekota2008chemical, yamanishi2011extracting, borgelt2002mining, coatney2003motifminer, brint1987algorithms} over molecule generation~\cite{jin2020hierarchical, hy2021multiresolution, jin_junction_2018, jin2020multi, guarino2017dipol} and interpretability of machine learning architectures~\cite{montavon_methods_2018,samek2021explaining,schnake2020higher,noutahi_towards_2020, mccloskey2019using, chen2020molecule, doi:10.1021/acs.jcim.0c01409, webel2020revealing, khasahmadi2019memory, 9810062} to coarse-graining~\cite{wang2019coarse, webb_graph-based_2019, chakraborty_encoding_2018}. However, to ensure the identification of meaningful moieties that can be utilized for a wide range of applications, a procedure is required (i) to be transferable w.r.t. molecule size, (ii) to provide a substructure decomposition of each molecule which preserves its respective global structure (required for, e.g., coarse-graining), and (iii) to allow for identifying several moieties of the same type in individual molecules (due to a common substructure often appearing multiple times). 
None of the methods mentioned above meets all of these criteria.

\begin{figure*}[!ht]
    \centering
    \includegraphics[width=\textwidth]{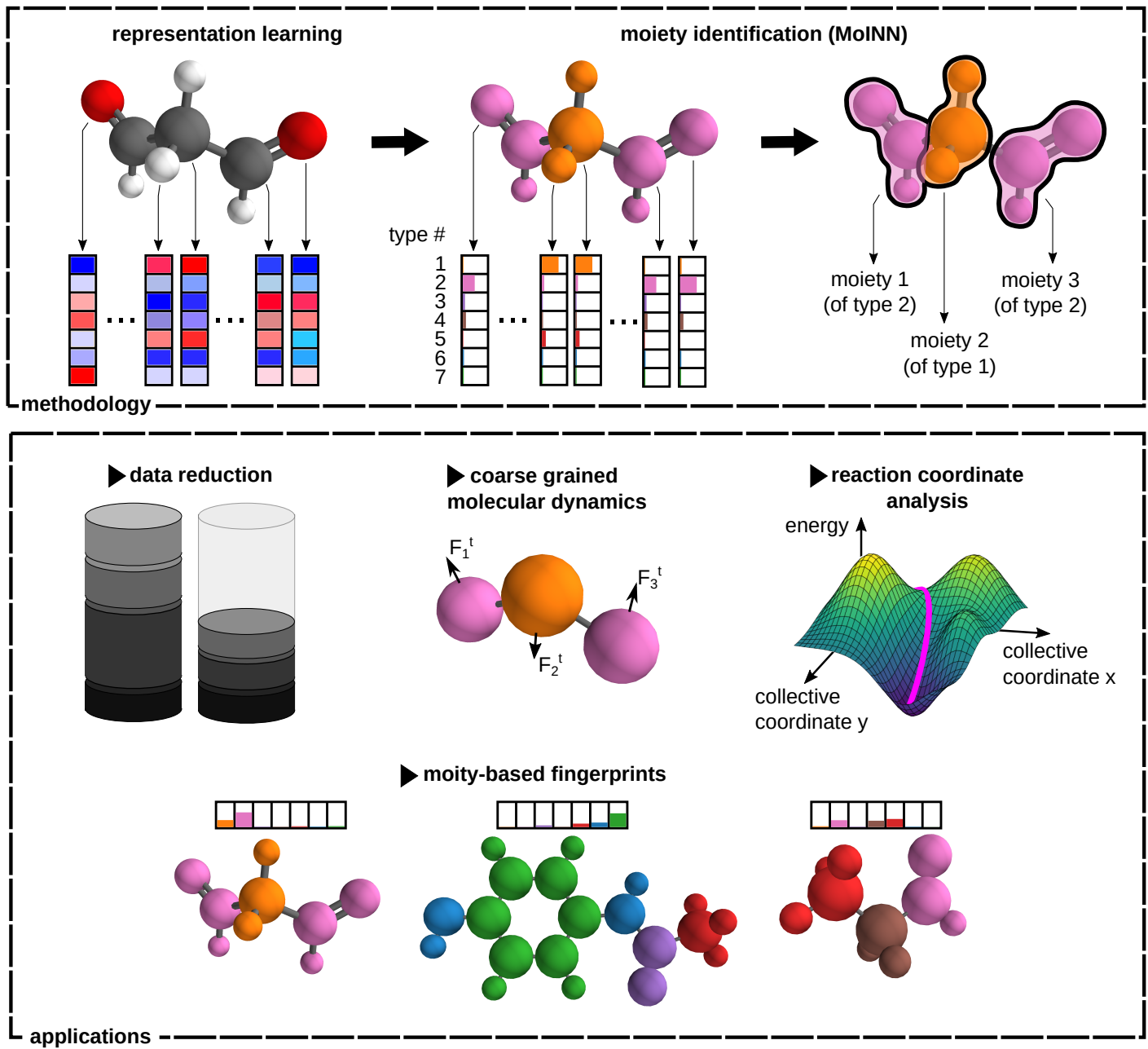}  
    \caption{\method{} methodology and applications. The \textbf{top} shows the moiety identification process for malondialdehyde. First, atomic feature representations (red and blue bars) are learned. Next, \method{} constructs type assignment vectors (pink and orange bars) based on these features. Each entry represents the probability of an atom to be assigned to a specific type of moiety (atoms are colored according to the highest atom-to-type affinity). Based on these assignments and the proximity of atoms, \method{} divides molecules into individual moieties. In this example, three chemical moieties of two distinct types associated with methylene (type 1, orange) and aldehyde (type 2, pink) groups are identified. The moiety representation allows for a variety of applications, which are shown on the \textbf{bottom}. They range from moiety-based fingerprint design, reaction coordinate analysis, and data reduction to coarse grained molecular dynamics.}
    \label{fig:pipeline}
\end{figure*}

In this work, we propose \method{} (Moiety Identification Neural Network) -- a method for the automatic identification of chemical moieties from the representations learned by MPNNs. This is achieved by constructing a soft assignment (or affinity) matrix from the atomic features, which maps individual atoms to different types of multi-atom substructures (Fig.~\ref{fig:pipeline}, top). By employing representations from MPNNs pretrained on molecular properties, the identified moieties are automatically adapted to the chemical characteristics of interest. Alternatively, it is possible to find chemically meaningful substructures by training MPNNs coupled with \method{} in an end-to-end manner. Here, only structural information is required and \textit{ab initio} calculations can be avoided. Crucially, \method{} is transferable between molecules of different sizes and automatically determines the appropriate number of moiety types. Multiple occurrences of the same structural motif within a molecule are recognized as the same type of moiety.

We demonstrate the versatility of \method{} by utilizing the identified chemical moieties to solve a range of tasks, which would otherwise require expert knowledge (Fig.~\ref{fig:pipeline}, bottom).
For example, the learned moiety types can serve as molecular fingerprints, which allow to estimate the properties of compounds from their composition, or used to extract the most representative entries from quantum chemical databases. Beyond that, moieties can be employed as coarse-grained representations of chemical structures, allowing to automatically determine \emph{beads} for the construction of coarse-grained force fields. Finally, we use \method{} to identify reaction coordinates in molecular trajectories based on the transformation of detected moieties.

\section{Method}\label{sec:method}

The automated identification of moieties with \method{} corresponds to a clustering of the molecule into different types of chemical environments. Hence, atoms in comparable environments, i.e.\ with similar feature representations (see Section~\ref{sec:MPNNs}), are likely to be assigned to the same cluster. In the following, the term ``environment types'' or short ``types'' will be used, since each cluster is associated with a specific substructure that exhibits particular chemical characteristics. Note that atoms belonging to the same environment type are not necessarily spatially close, because similar substructures may appear multiple times at distant locations in a molecule. This is why, after atoms have been assigned to environment types (see Section~\ref{sec:type-based}), individual (spatially disconnected) chemical moieties can be found by introducing an additional distance criterion (see Section~\ref{sec:ind_moi}). Both steps are combined to arrive at an unsupervised learning objective for decomposing molecules into chemical moieties (see Section~\ref{sec:opt_prob}).

\subsection{Representation Learning in Message Passing Neural Networks}\label{sec:MPNNs}

Message passing neural networks (MPNNs)~\cite{gilmer_neural_2017} are able to learn atomic feature representations from data in an end-to-end manner (without relying on handcrafted features). They achieve state-of-the-art performance for molecular property prediction, solely taking atomic numbers and atom positions as inputs~\cite{schutt_quantum-chemical_2017, schutt_schnet_2018, unke2019physnet,klicpera2020directional,batzner2021se,schutt2021equivariant,unke2021spookynet}. The representation learning scheme of an MPNN can be described as follows. First, atomic features are initialized to embeddings based on their respective atomic numbers (all atoms of the same element start with the same representation). Subsequently, the features of each atom are iteratively updated by exchanging ``messages'' with neighboring atoms, which depend on their current feature representations and distances. After several iterations, the features encode the relevant information about the chemical environment of each atom. In this work, we use SchNet~\cite{schutt_schnet_2018} to construct atomic feature representations. In general, however, \method{} is applicable to any other representation learning scheme.

\subsection{Assigning Atoms to Environment Types}\label{sec:type-based}

Starting from $F$-dimensional atomic feature representations $\mathbf{x}_1,\dots,\mathbf{x}_N$ of $N$ atoms (e.g.\ obtained from an MPNN), a type assignment matrix $\mathbf{S}$, which maps individual atoms to different environment types, is constructed. Following a similar scheme as Bianchi~et.~at.~\cite{bianchi_mincut_2019}, the type assignment matrix is given by
\begin{equation}
    \mathbf{S} = \mathrm{softmax}\left(
    \mathrm{SiLU}\left(\mathbf{XW}_1\right)\mathbf{W}_2
    \right)~,
\label{eq:type_ass}
\end{equation}
where ${\mathbf{W}_1\in\mathbb{R}^{F\times K}}$ and ${\mathbf{W}_2\in\mathbb{R}^{K\times K}}$ are trainable weight matrices, the $n$-th row of the feature matrix ${\mathbf{X}\in\mathbb{R}^{N\times F}}$ is the representation $\mathbf{x}_n \in\mathbb{R}^{F}$ of atom~$n$, and SiLU is the Sigmoid Linear Unit activation function~\cite{hendrycks2016gaussian}. Here, $K$ is a hyperparameter that denotes the maximum number of possible types. As will be shown later, a meaningful number of types is automatically determined from data and largely independent of the choice of $K$ (see Section~\ref{sec:opt_prob}). The $\mathrm{softmax}$ function ensures that entries $\mathrm{S}_{nk}$ of
the ${N\times K}$ matrix $\mathbf{S}$ obey ${\sum_k \mathrm{S}_{nk}=1~\forall n}$ with ${\mathrm{S}_{nk}>0}$. Thus, each row of $\mathbf{S}$ represents a probability distribution over the $K$ environment types, with each entry $\mathrm{S}_{nk}$ expressing how likely atom~$n$ should be assigned to type~$k$. Even though assignments are ``soft'', i.e.\ every atom is partially assigned to multiple environment types, the $\mathrm{softmax}$ function makes it unlikely that more than one entry in each row is dominant (closest to~$1$). The advantage of a soft type assignment matrix is that its computation is well suited for gradient-based optimization. In other contexts, however, it might be more natural to assign atoms unambiguously to only one environment type. For this reason, we also define a ``hard'' type assignment matrix $\mathbf{S}_\mathrm{h} \in \mathbb{R}^{N\times K}$ with entries
\begin{equation}
    \mathrm{S}_{\mathrm{h}, nk}=    
    \begin{cases}
      1 & \mathrm{S}_{nk}> \mathrm{S}_{nj}~\forall j \in [0,K)\backslash\{k\} \\
      0 & \mathrm{otherwise}
    \end{cases}~,
    \label{eq:type_ass_hard}
\end{equation}
such that each row contains exactly one non-zero entry equal to~$1$.

The atomic feature representations making up the matrix~$\mathbf{X}$ can either be provided by a pretrained model, or learned in an end-to-end fashion. Depending on the use case, both approaches offer their respective advantages: Since the type assignment matrix $\mathbf{S}$ is directly connected to $\mathbf{X}$ via Eq.~\ref{eq:type_ass}, pretrained features allow to find types adapted to a specific property of interest. End-to-end learned representations have the advantage that they do not rely on any reference data obtained from computationally demanding quantum mechanical calculations. Instead, they are found from structural information by optimizing an unsupervised learning problem (see Section~\ref{sec:opt_prob}).

\subsection{Assigning Atoms to Individual Moieties}\label{sec:ind_moi}

Molecules may consist of multiple similar or even identical substructures. Consequently, distant atoms with comparable local environments can be assigned to the same type, even though they do not necessarily belong to the same moiety (see Fig.~\ref{fig:pipeline}). To find the actual chemical moieties, i.e.\ groups of nearby atoms making up a structural motif, we introduce the $N \times N$ moiety similarity matrix given by
\begin{equation}
    \mathbf{C} =  \mathbf{SS}^T \circ \mathbf{A}~, \label{eq:cluster-similarity}
\end{equation}
where ``$\circ$'' denotes the Hadamard (element-wise) product.
Here, the $N\times N$ matrix $\mathbf{SS}^T$ measures the similarity of the type assignments between atoms, i.e.\ its entries are close to~$1$ when a pair of atoms is assigned to the same environment type and close to~$0$ otherwise. The adjacency matrix ${\mathbf{A}\in [0,1]^{N\times N}}$ on the other hand captures the proximity of atoms. Its entries are defined as
\begin{equation}
\mathrm{A}_{ij}(r_{ij}) =
\begin{cases}
    0.5 \left(1 + \cos\left(\frac{\pi r_{ij}}{r_\mathrm{cut}}\right)\right)
      & r_{ij} < r_\mathrm{cut} \\
    0 & r_{ij} \geqslant r_\mathrm{cut} 
\end{cases} \,,
\label{eq:adjacency_entries}
\end{equation}
where $r_{ij}$ is the pairwise distance between atoms~$i$~and~$j$ and $r_\mathrm{cut}$ is a cutoff distance. For simplicity, we employ a cosine cutoff to assign proximity scores, but more sophisticated schemes are possible (e.g. based on the covalent radii of atoms). The combination of $\mathbf{SS}^T$ and $\mathbf{A}$ ensures that the entries of the similarity matrix $\mathbf{C}$ are close to 1 only if two atoms are both assigned to the same type \emph{and} spatially close, in which case they belong to the same chemical moiety.

Analogous to the hard assignment matrix $\mathbf{S}_\mathrm{h}$ (see Eq.~\ref{eq:type_ass_hard}), a hard moiety similarity matrix $\mathbf{C}_\mathrm{h}$, which unambiguously assigns atoms to a specific moiety, might be preferable over Eq.~\ref{eq:cluster-similarity} in some contexts. To this end, we define the matrix
\begin{equation}
    \mathbf{C}_\mathrm{h}^0 =  \mathbf{S}_\mathrm{h}\mathbf{S}_\mathrm{h}^T \circ \mathbf{A}_\mathrm{cov}~.
\end{equation}
where $\mathbf{A}_\mathrm{cov}$ has entries of 1 for each atom-pair connected by a covalent bond (see Section~S1\dag) and 0 otherwise. $\mathbf{C}_\mathrm{h}^0$ describes a graph on which breadth-first search~\cite{10.5555/1410219} is performed to find its connected components (moieties). This yields a hard similarity matrix $\mathbf{C}_\mathrm{h}$, which maps atoms unambiguously to their individual moieties (for further details, please refer to Section~S2\dag).

\subsection{Optimization of Environment Type Assignments and Moiety Assignments}\label{sec:opt_prob}

Chemical moieties are identified by minimizing the unsupervised loss function
\begin{equation}
\mathcal{L} = \mathcal{L}_\mathrm{cut} + \mathcal{L}_\mathrm{ortho} + \alpha\mathcal{L}_\mathrm{ent},
\label{eq:unsup_loss}
\end{equation}
where $\mathcal{L}_\mathrm{cut}$, $\mathcal{L}_\mathrm{ortho}$, and $\mathcal{L}_\mathrm{ent}$ are cut loss, orthogonality loss, and entropy loss, and $\alpha$ is a trade-off hyperparameter.
The cut loss $\mathcal{L}_\mathrm{cut}$~\cite{bianchi_mincut_2019} penalizes ``cutting'' the molecule, i.e.\ assigning spatially close atoms to different moieties. It is defined as
\begin{align*} 
\mathcal{L}_\mathrm{cut} &=
    -\frac
        {Tr\left(\mathbf{C}^T\tilde{\mathbf{A}}\mathbf{C}\right)}
        {Tr\left(\mathbf{C}^T\tilde{\mathbf{D}}\mathbf{C}\right)},
\end{align*}
where $\tilde{\mathbf{A}}=\mathbf{D}^{-1/2}\mathbf{AD}^{-1/2}\in\mathbb{R}^{N\times N}$ is a symmetrically normalized adjacency matrix (see Eq.~\ref{eq:adjacency_entries}). The degree matrix $\mathbf{D}\in\mathbb{R}^{N\times N}$ is diagonal with elements $\mathrm{D}_{ii} = \sum_j^N \mathrm{A}_{ij}$, where $\mathrm{A}_{ij}$ are the entries of $\mathbf{A}$. Consequently, $\tilde{\mathbf{D}}$ is the degree matrix obtained from the entries of $\tilde{\mathbf{A}}$. 

To avoid converging to the trivial minimum of $\mathcal{L}_\text{cut}$ where all atoms are assigned to the same moiety and type, the orthogonality loss~\cite{bianchi_mincut_2019}
\begin{align*} 
\mathcal{L}_\mathrm{ortho} &=
    \left\Vert
        \frac{\mathbf{SS}^T}{\left\Vert\mathbf{SS}^T\right\Vert_F} - 
        \frac{\mathbf{I}_N}{\sqrt{N}}
    \right\Vert_F 
\end{align*} 
drives the type assignment vectors of different atoms (i.e.,\ the rows of $\mathbf{S}$) to be (close to) orthogonal. Here, $\mathbf{I}_N$ is the $N\times N$ identity matrix and $\Vert \cdot \Vert_F$ is the Frobenius norm.

Finally, the entropy term~\cite{ying_hierarchical_2018}
\begin{align*} 
\mathcal{L}_\mathrm{ent} &= 
    - \frac{1}{N}\sum_{nk} \mathrm{S}_{nk} \ln{(\mathrm{S}_{nk})}
\end{align*}
favors ``hard'' assignments and indirectly limits the number of used types (here, $\mathrm{S}_{nk}$ are the entries of $\mathbf{S}$, see Eq.~\ref{eq:type_ass}). Without this term, there is no incentive to use fewer than $K$ types, i.e.,\ the model would eventually converge to use as many different types as possible. Hence, by introducing the entropy term, we avoid relying on expert knowledge for choosing $K$ and instead facilitate learning a meaningful number of types from data.

In principle, the number of used types still depends on $K$ and the entropy trade-off factor $\alpha$. However, as is shown in Section~S3\dag, there is a regime of $\alpha$ where the number of used types is largely independent of $K$ (as long as $K$ is sufficiently large). Hence, we arbitrarily choose $K=100$ in our experiments if not specified otherwise.

\section{Applications}\label{sec:experiments}

This section describes several applications of \method{}. First, we use \method{} to identify common moieties in molecular data (Section~\ref{cl_qm9}). Leveraging these insights, we select representative examples from a database of structures to efficiently reduce the number of reference calculations required for property prediction tasks (Section~\ref{data_red}). Next, an automated pipeline for coarse-grained molecular dynamics simulations built on top of \method{} is described (Section~\ref{cg-md}). Finally, we demonstrate how to utilize \method{} for automatically detecting reaction coordinates in molecular dynamics trajectories (Section~\ref{dyn_clus}).

\subsection{Identification of Chemical Moieties}\label{cl_qm9}
    
To demonstrate the automatic identification of chemical moieties, we apply \method{} to the QM9 dataset~\cite{ramakrishnan2014quantum}. Here, two different models are considered: One utilizes fixed feature representations provided by a SchNet model pretrained to predict energies, while the other model is trained in an end-to-end fashion on purely structural information. In the following, these will be referred to as the pretrained model and the end-to-end model, respectively. Details on the training of SchNet and \method{}, as well as, a comparison between pretrained model and end-to-end model can be found in Section~S4\dag. \revision{Also the impact of varying training data size on \method{} outputs is shown there.}
    
\begin{figure}[tb]
    \centering
    \includegraphics[width=0.49\textwidth]{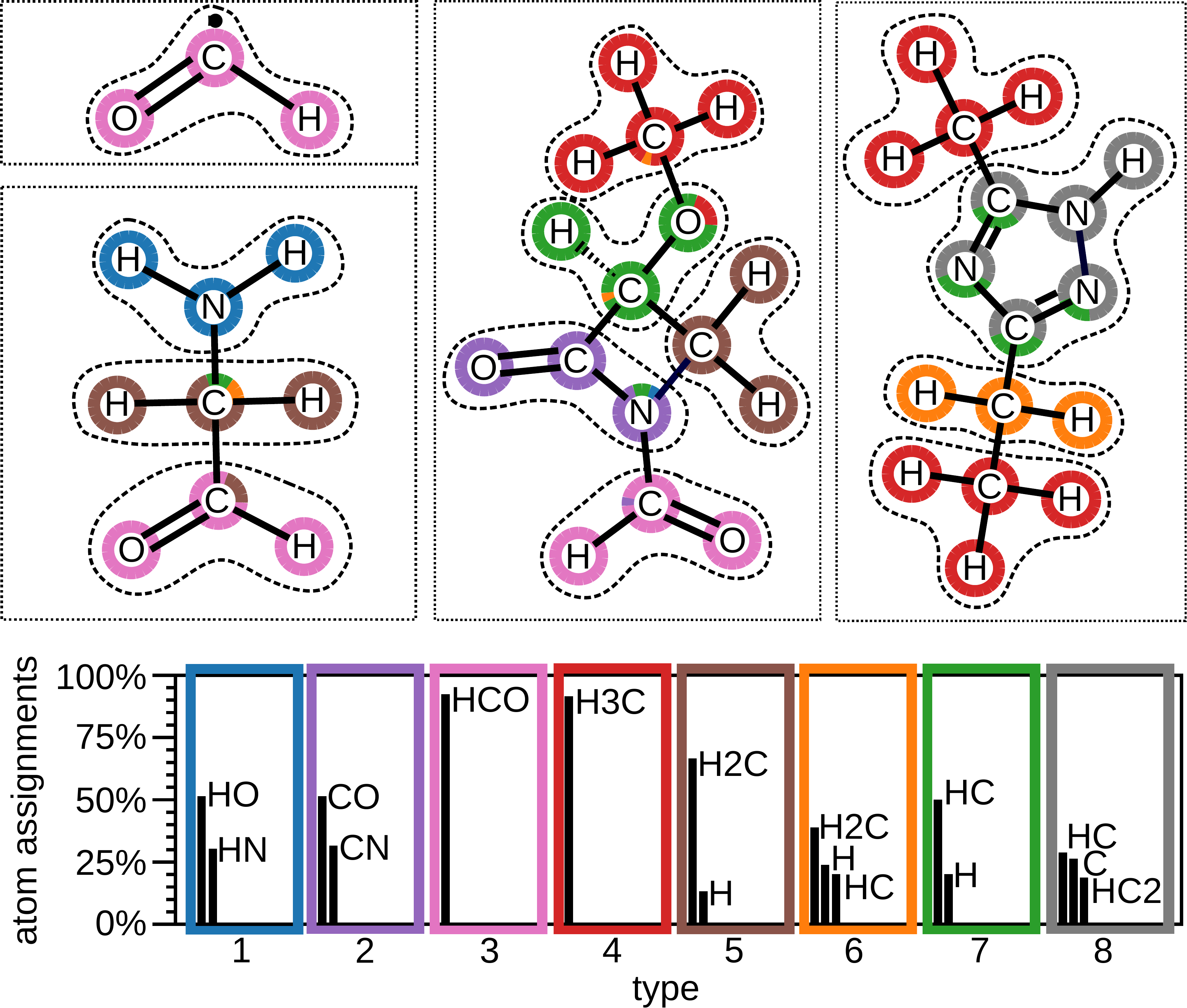}
    \caption{Common moieties of the QM9 dataset. The \textbf{top} shows four exemplary molecules along with type assignments (colored circles) and moieties (enclosed by dashed lines). The \textbf{bottom} shows the distribution of environment types and corresponding most common moieties for the test set (1000 molecules), black bars indicate the relative amount of atoms assigned to the respective moieties. For each environment type, over $70\%$ of its atom assignments correspond to at most three different moieties.}
    \label{fig:common_substruc}
\end{figure}
Figure~\ref{fig:common_substruc} depicts the results for the pretrained \method{} model evaluated on a test set of 1,000 molecules that were excluded from the training procedure. The top shows four exemplary molecules with corresponding type assignments and moieties. As expected, we observe that moieties of the same type may occur across different molecules, as well as multiple times in a single molecule. The evaluation of environment types and corresponding moieties for all 1,000 molecules (see bottom of Fig.~\ref{fig:common_substruc}) shows that each type is associated with a small set of similar moieties, i.e., the environment types form a ``basis'' of common substructures that can be combined to form all molecules contained in the dataset. \revision{In Section~S5.1\dag, we evaluate \method{} w. r. t. various ring systems and the largest identified moieties. We observe that while saturated rings are predominantly divided into several small moieties, MoINN tends to identify aromatic rings as individual entities.}
    
To verify that the type assignments are chemically meaningful, we use them to construct molecular fingerprints, from which different chemical properties are predicted via \revision{a linear regression model (for more details refer to Section~S5.1\dag)}. Although it is unrealistic to expect state-of-the-art performance with such a simple model, meaningful molecular fingerprints should at least perform on-par with handcrafted variants. 
The type-based fingerprints are constructed from the assignments learned by the end-to-end \method{} model as
\begin{equation}
    \mathbf{h}_\mathrm{\method{}} =  \sum_n \mathrm{S}_{nk}(\mathbf{X}) ~,
\label{eq:fingerprints}
\end{equation}
where $\mathbf{X}$ denotes the feature matrix and $\mathrm{S}_{nk}$ is the assignment matrix entry for the $n$-th atom and the $k$-th type. \revision{The feature size of the type-based fingerprints is given by the number of environment types $K=100$. However, due to the sparsity of the environment types, the effective number of features is $17$ (see also Section~S4\dag). For comparison, we use handcrafted fragmentation of molecules stated by Nannolal~et.~al.~\cite{NANNOOLAL2008117}, fingerprints provided by the fragment catalogue and fragment generator of RDKit~\cite{rdkit}, and Morgan fingerprints~\cite{rogers2010extended} with feature sizes $86$, $15387$ and $100$, respectively.} 
Table~\ref{tab:lin_model} compares the \revision{test errors of the corresponding linear models}.

The type-based fingerprints significantly outperform \revision{the other considered fingerprints}, which suggests that the environment types provided by \method{} are a chemically meaningful representation of the molecules in the dataset.
\revision{The performance is particularly good for the considered extensive properties $U_0$, $H$, and $F$. The reason for this is that, in contrast to the other fingerprints, information about the molecule size is implicitly contained in the fingerprints provided by \method{}.}
\begin{table}[h]
\small
  \caption{\ Mean absolute error of the predicted dipole moment $\mu$, internal energy $U_0$, the enthalpy $H$, and the free energy $F$ based on linear regression on different molecular fingerprints.}
  \label{tbl:example1}
  \begin{tabular*}{0.48\textwidth}{@{\extracolsep{\fill}}lllll}
    \hline
    property & \method{} & \thead{Nannoolal\\et.~al.~\cite{NANNOOLAL2008117}}  & RDKit~\cite{rdkit} & \revision{Morgan~\cite{rogers2010extended}}\\
    \hline
    $\mu$ (Debye) & $\mathbf{0.07}$ & $0.89$ & $0.62$ & \revision{$0.19$}\\
    $U_0$ (eV) & $\mathbf{1.46}$ & $454.66$ & $303.44$ & \revision{$513.78$}\\
    $H$ (eV) & $\mathbf{1.64}$ & $455.04$ & $318.25$ & \revision{$512.65$}\\
    $F$ (eV) & $\mathbf{0.65}$ & $464.82$ & $320.85$ & \revision{$518.61$}\\
    \hline
  \end{tabular*}
\label{tab:lin_model}
\end{table}

\subsection{Sampling of Representative Molecules}\label{data_red}

\begin{figure}[tb]
    \centering
    \includegraphics[width=0.44\textwidth]{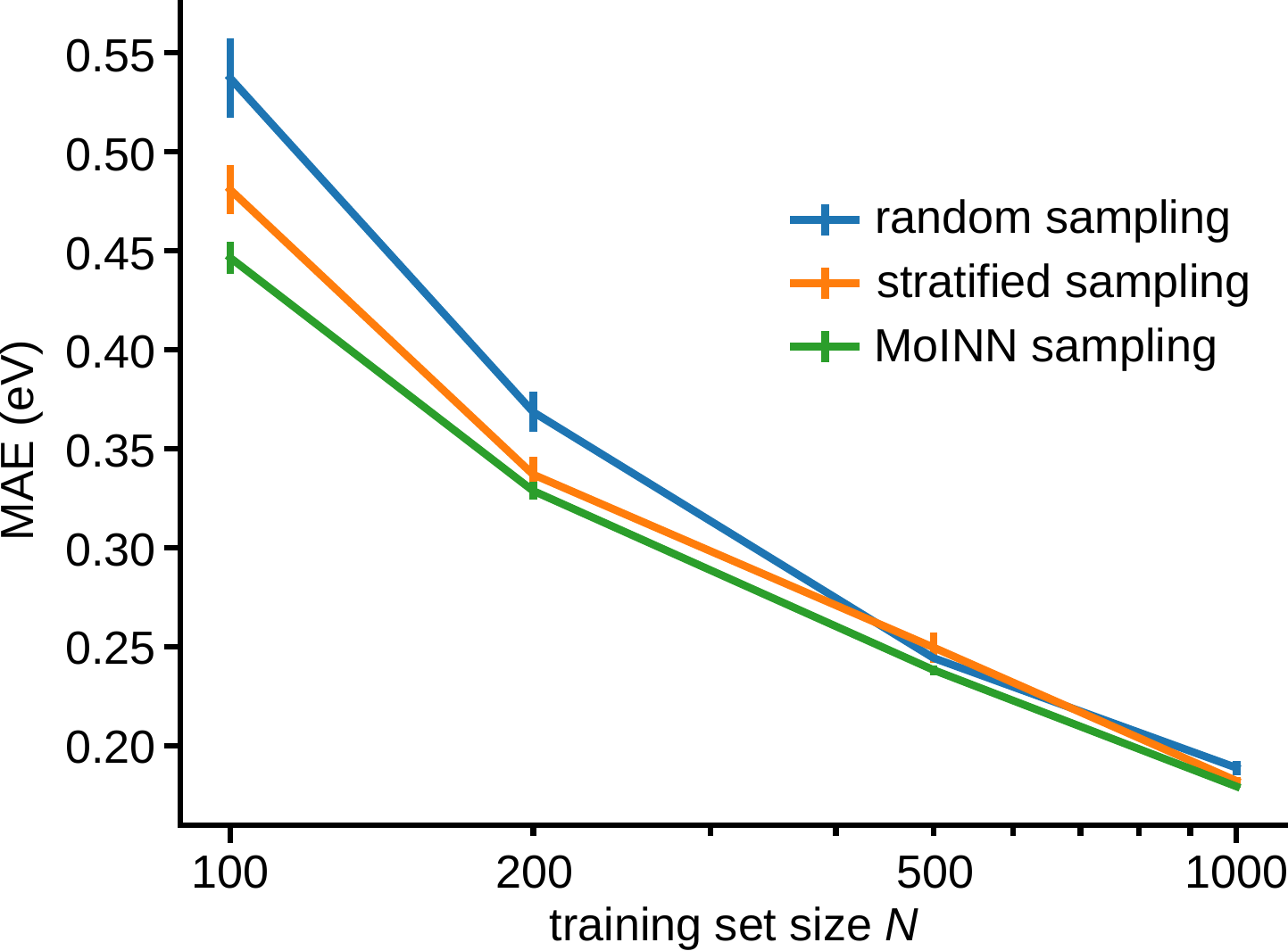}
    \caption{Mean absolute error (MAE) of energy predictions for SchNet models trained on randomly sampled training sets (blue), \revision{training sets obtained by stratified sampling (orange)} and training sets selected with \method{} (green). Each data point is averaged over \revision{five} independent training runs and standard errors are indicated by error bars.}
    \label{fig:reduced_data}
\end{figure}
    
The quality of the reference dataset used to train ML models greatly impacts their generalization performance~\cite{vazquez2021impact}. Since the calculation of molecular properties at high levels of theory is computationally demanding, it is desirable to find ways to reduce the amount of reference data needed for training accurate machine learning models. One way to achieve this is by sampling a representative subset of datapoints from chemical space (instead of choosing points randomly). Here, we employ the type-based fingerprints (Eq.~\ref{eq:fingerprints}) described in the previous section to find a subset of molecules as small as possible, which still represents the QM9 dataset sufficiently well. To this end, we minimize the loss function
\begin{equation}
    \mathcal{L}_\mathrm{data} = \left\Vert \mathbf{W}\mathbf{H}_\mathrm{\method{}} - \mathbf{H}_\mathrm{\method{}} \right\Vert_F + \lambda \sum_j \sqrt{ \sum_i w_{ij}^2}~.
\label{eq:data}
\end{equation}
${\mathbf{H}_\mathrm{\method{}}\in\mathbb{R}^{D\times K}}$ denotes the fingerprint matrix of $D$~molecules, where each row is given by the fingerprint vector $\mathbf{h}_\mathrm{\method{}}$ of a specific molecule. $\mathbf{W} \in\mathbb{R}^{D\times D}$ is a trainable weight matrix with entries $\{w_{ij}\}$. The first term in Eq.~\ref{eq:data} describes the reconstruction error. To avoid converging to the trivial solution, where the trainable matrix $\mathbf{W}$ is simply the identity matrix, we introduce a regularization term that enforces sparse rows in the weight matrix $\mathbf{W}$. The trade-off between both terms can be tuned by the factor $\lambda$, i.e.\ larger values of $\lambda$ will select a smaller subset of representative molecules. Intuitively, minimizing Eq.~\ref{eq:data} corresponds to selecting a small number of molecules as ``basis vectors'', from which all other molecules can be (approximately) reconstructed by linear combination.
    
Based on this procedure, we select \revision{several} QM9 subsets of different size as training sets and compare them to randomly sampled subsets, \revision{and subsets obtained by stratified sampling w. r. t. the number of atoms in each molecule.} For each of these subsets, we train \revision{five} SchNet models and evaluate their average performance (Fig.~\ref{fig:reduced_data}). Models trained on subsets chosen by \method{} perform significantly better than those trained on randomly sampled subsets and stratified sampled subsets. This effect is most pronounced for small training set sizes. \revision{Thus, selecting data with \method{} is most useful in a setting where only few data points can be afforded, e.g.\ when using a high level of theory to perform reference calculations. For more details on the experiment and a comprehensive discussion of the results please refer to Section~S5.2\dag.}

\subsection{Coarse-Grained Molecular Dynamics}\label{cg-md}

While ML force fields accelerate \textit{ab initio} MD simulations by multiple orders of magnitude~\cite{unke2021machine}, the study of very large molecular structures is still computationally demanding. Coarse-graining~(CG) reduces the dimensionality of the problem by representing groups of atoms as single interaction sites. Most approaches rely on systematically parametrized CG force fields~\cite{marrink2007martini, brini_systematic_2013}, but also data driven approaches have been proposed~\cite{husic2020coarse, wang_ensemble_2020, wang_machine_2018, zhang_deepcg:_2018}. In both cases, however, the coarse-grained ``beads'' are usually determined manually by human experts~\cite{riniker2012developing}.


Here, we propose an automated pipeline for coarse-grained molecular dynamics simulations (CG-MD), which comprises atomistic SchNet models for noise reduction, \method{} for reducing the molecule's degrees of freedom, and a SchNet model trained on the CG representation for simulating the CG dynamics. 
We apply this approach to the trajectory of alanine-dipeptide in water~\cite{nuske2017markov, wehmeyer2018time}, which is a commonly used model system
for comparing different CG methods~\cite{wang_machine_2018, wang_ensemble_2020, husic2020coarse}. 
    
\begin{figure}[tb]
    \centering
    \includegraphics[width=0.41\textwidth]{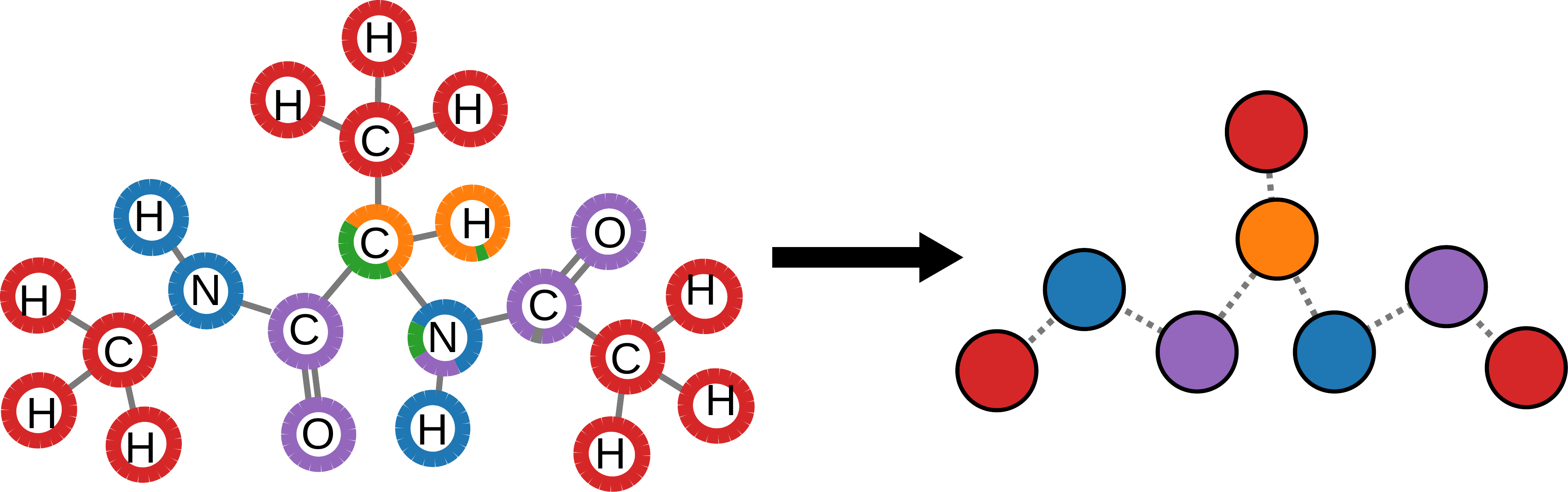}
    \caption{Automated coarse-graining with \method{}. On the \textbf{left}, the alanine-dipeptide molecule is depicted at atomic resolution, assigned environment types are indicated by colored circles. On the \textbf{right}, the corresponding coarse-grained representation, derived from environment types and moiety assignments, is shown.}
\label{fig:alanine_CG}
\end{figure}

The CG representation, shown in Fig.~\ref{fig:alanine_CG}, is inferred from the environment types and moieties provided by the pretrained \method{} model described in Section~\ref{cl_qm9}, which has been trained on the QM9 dataset. \revision{For a comparison to conventional CG representations such as, e.g., OPLS-UA\cite{doi:10.1021/ja00214a001, doi:10.1021/ja9621760}, or an automated CG approach\cite{potter2021automated} for the Martini force field~\cite{marrink2007martini}, please refer to section~S5.3\dag.} 
The original atomistic trajectory of alanine-dipeptide does not include reference energies. 
This is because the dynamics have been simulated in solvent, which introduces noise to the energy of the system if the solvent is not modeled explicitly. The data contains forces for all atoms in the alanine-dipeptide molecule, which implicitly include interactions with solvent molecules. However, sparsely sampled transition regions between conformers are challenging to learn with force targets only. 
Coarse-graining introduces additional noise on the energies and forces~\cite{wang_machine_2018} since some information about the atom positions is discarded.

To reduce the noise, we train an ensemble of five SchNet models to provide a force field for alanine-dipeptide at atomic resolution. Subsequently, we use the corresponding forces $\hat{\mathbf{F}}$ and energies $\hat{U}$ as targets for training the CG SchNet model in a force-matching scheme adapted from Refs.~\citenum{wang_machine_2018, wang_ensemble_2020, husic2020coarse, noid_multiscale_2008, noid_multiscale_2008-1} (see Section~S5\dag\ for details). 
\begin{figure}[tb]
    \centering
    \includegraphics[width=0.41\textwidth]{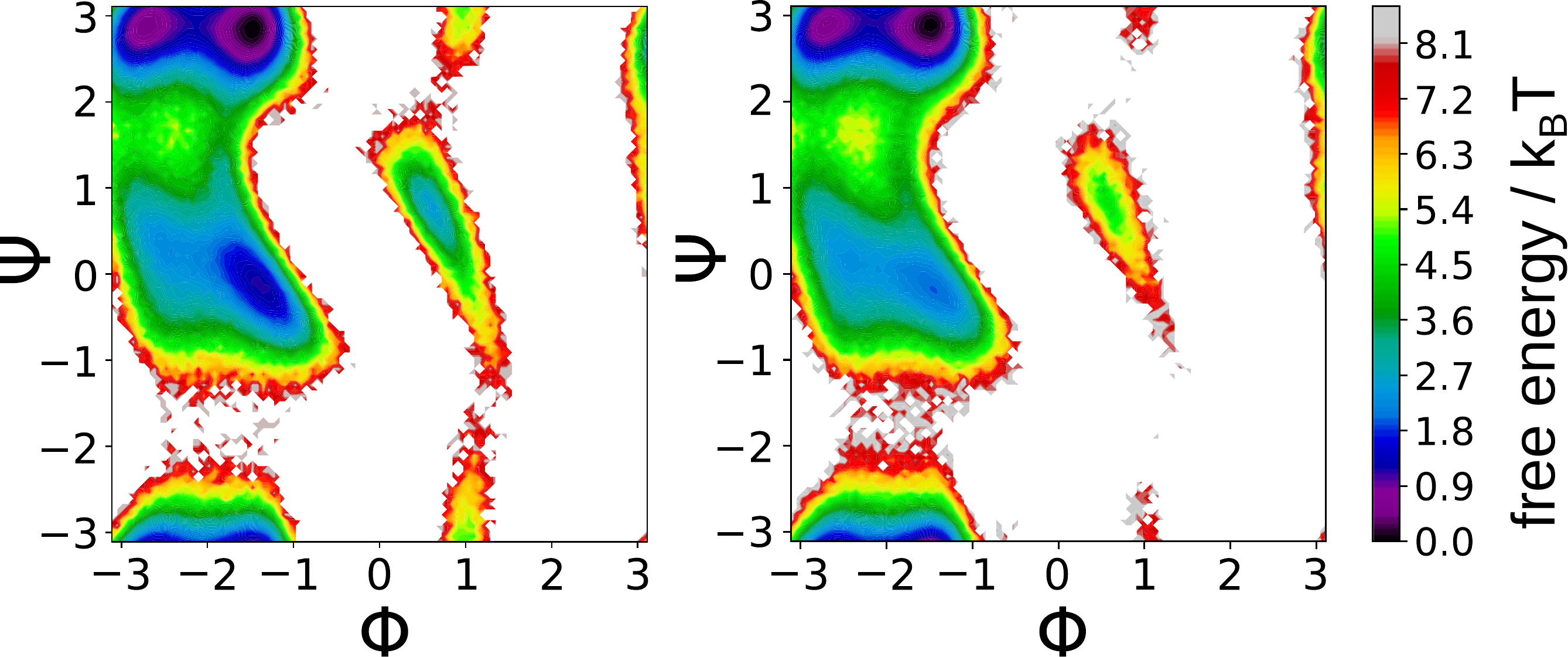}
    \caption{Ramachandran plots of the free energy surface of alanine-dipeptide with respect to the torsion angles $\phi$ and $\psi$ for the atomistic MD ({\textbf{left}}) and the coarse-grained MD ({\textbf{right}}) ($\phi$ and $\psi$ are computed with respect to the coarse-grained geometry).}
    \label{fig:ramachandran}
\end{figure}
Based on the CG SchNet model, we run MD simulations in the NVT ensemble at room temperature (\SI{300}{\kelvin}). The trajectories are initialized according to the Boltzmann distribution at the six minima of the potential energy surface. For keeping the temperature constant, we use a Langevin thermostat. Fig.~\ref{fig:ramachandran} shows that the free energy surfaces derived from the all-atom and CG trajectories are in good agreement.

\begin{figure*}[tb]
    \centering
    \includegraphics[width=0.65\textwidth]{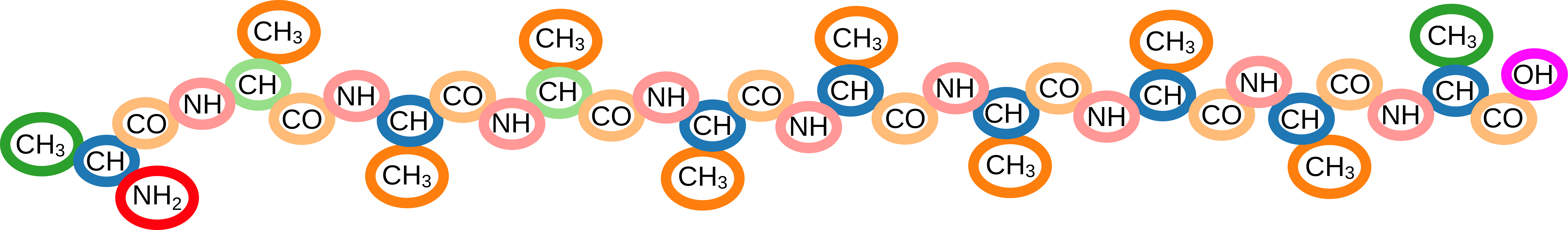}
    \caption{\revision{CG-representation of deca-alanine inferred from environment types provided by \method{}.}}
\label{fig:decaalanine_CG}
\end{figure*}

\revision{
\method{} also allows for coarse-graining structures outside the scope of QM9. This is shown in Fig.~\ref{fig:decaalanine_CG} for decaalanine. The provided CG representation resembles the commonly used OPLS-UA representation~\cite{doi:10.1021/ja00214a001, doi:10.1021/ja9621760}. However it is striking that the type of terminal methyl groups differs from that of the backbone methyl groups, while in the OPLS-UA representation, by construction, those groups are considered to be interchangeable. For more details how the CG representation is derived from the provided environment types, see Section~S5.3\dag.
}

\subsection{Dynamic Clustering and Reaction Coordinate Analysis}\label{dyn_clus}

\method{} is also capable of learning environment types for molecular trajectories. In this case, the types describe ``dynamic clusters'', which can be useful, e.g., for determining collective variables that describe chemical reactions. As a demonstration of this concept, we consider two chemical reactions, namely the proton transfer reaction in malondialdehyde and the Claisen rearrangement of allyl {\it p}-tolyl ether (see Refs.~\citenum{schutt2019unifying}
~and~\citenum{gastegger2021machine} for details on how the trajectories were obtained). We train individual end-to-end \method{} models on each reaction trajectory. 

\begin{figure}[tb]
    \centering
    \includegraphics[width=0.49\textwidth]{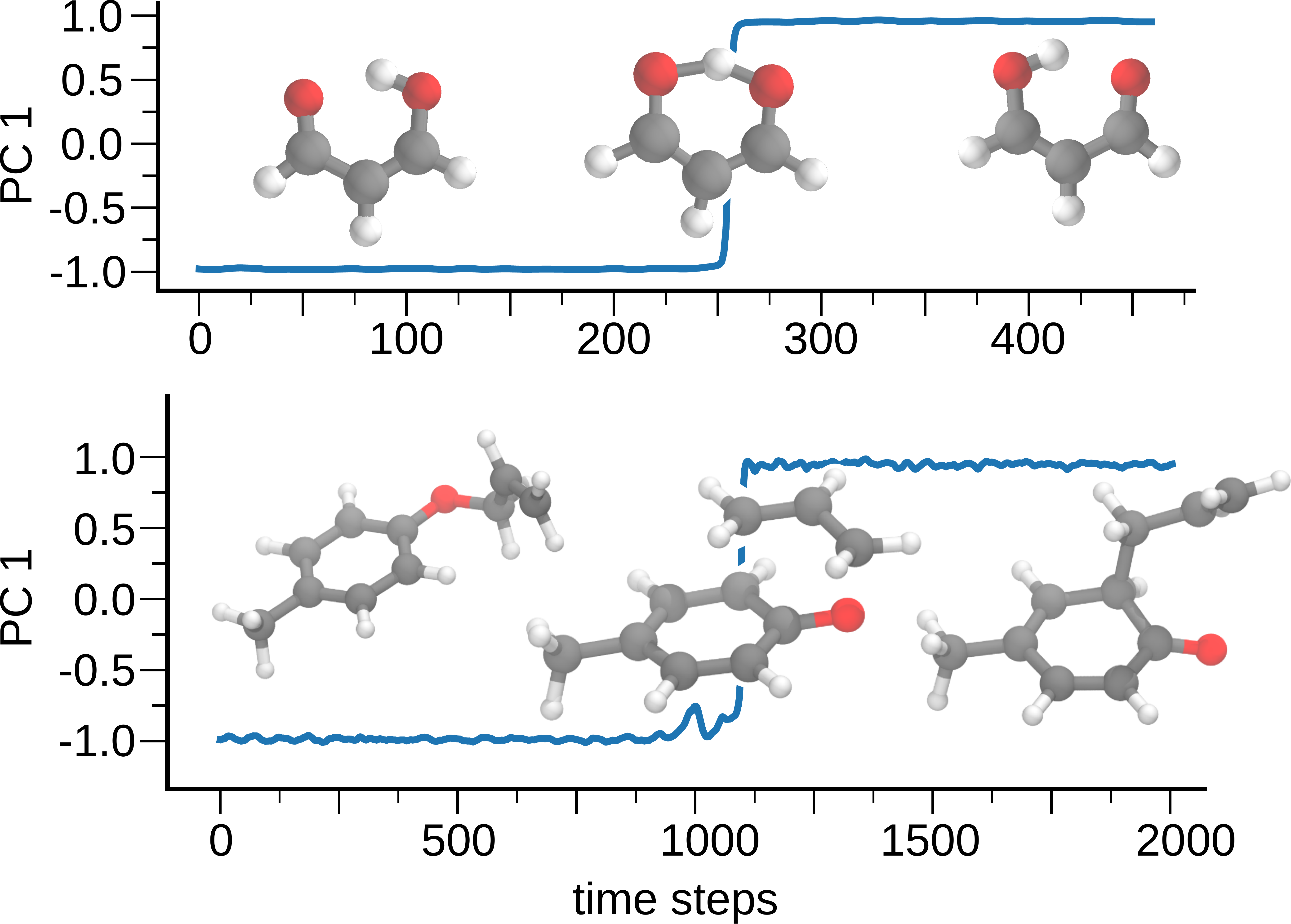}
    \caption{Automatic detection of reaction coordiantes for the proton transfer in malondialdehyde (\textbf{top}) an the Claisen rearrangement of allyl {\it p}-tolyl ether (\textbf{bottom}). The identified reaction coordinate is plotted w.r.t the time step of the respective trajectory.}
    \label{fig:reaction_coordinate}
\end{figure}

For each time step in the trajectory, we construct a high-dimensional coordinate vector
\begin{equation*}
    \mathbf{h}_\mathrm{dyn} = (S_{11}, S_{12}, ..., S_{1K}, S_{21}, ..., S_{2K}, ... ,S_{NK})~,
\end{equation*}
which consists of the type assignment matrix entries $\{S_{nk}\}$. By applying standard dimensionality reduction methods like principal component analysis (PCA)~\cite{hotelling1933analysis, pearson1901liii}, it is possible to extract a low-dimensional representation of the largest structural changes in the trajectory. For the two considered cases, we find that a one-dimensional reaction coordinate given by the first principal component provides a good description of the dynamic process and shows a prominent ``jump'' when the reaction happens (see Fig.~\ref{fig:reaction_coordinate}). 
\revision{In Section~S5.4\dag, we show that the reaction coordinate derived from MoINN allows for a more clear distinction between reactants and product than simply using the adjacency matrix based on the pairwise distances of atoms.}
    
\section{Discussion}
Owing to its computational efficiency, interpretability, and transferability, 
\method{} is applicable to a wide range of different tasks which otherwise rely on expert knowledge. \method{} involves a representation-based pooling approach which shares common ideas with the graph-pooling approaches DiffPool~\cite{ying_hierarchical_2018} and MinCut~\cite{bianchi_mincut_2019}. Latter describe the acquisition of a soft assignment matrix, which allows to pool graph nodes (atoms) to representative super-nodes (atom groups). In DiffPool the assignment matrix is learned utilizing GNNs, while MinCut employs a multilayer perceptron (MLP) architecture. Both methods introduce unsupervised loss functions to ensure a reasonable number of super-nodes while preferably grouping nearby nodes. Similar to MinCut, \method{} learns a mapping from atomic feature representations to type assignments by two dense layers (an MLP). The shallow network architecture results in computationally cheap training and inference. As the most prominent difference, \method{} distinguishes between individual moieties and environment types, while DiffPool and MinCut handle those entities interchangeably. As a result, \method{} stands out with regards to interpretability and transferability.

The distinction between moieties and environment types allows for a more detailed analysis of the identified substructures. While the environment types explain the composition and conformation of the molecular substructures, the moieties represent individual molecular building blocks. Besides the benefits with regards to interpretability, the distinction between moieties and environment types allows to identify multiple identical moieties in the same molecule. This feature is particularly useful for molecular systems since those often possess atom groups (moieties) of the same type multiple times. In contrast, pooling distant nodes is penalized when utilizing MinCut or DiffPool. Hence, even though some distant nodes might exhibit similar feature representations, they are unlikely to be grouped together. This makes it difficult to find common moieties and might hamper transferability w.r.t. molecules of different size, since the mapping between atoms and atom groups becomes more sensitive to small changes in the feature representations. For more details on this problem, please refer to Section~S6\dag.

\section{Conclusion and Outlook}\label{sec:outlook}
We have introduced \method{}, a versatile approach capable of automatically identifying chemical moieties in molecular data from the representations learned by MPNNs. Depending on the task at hand, \method{} can be trained based on pretrained representations or in an end-to-end fashion. While pretrained representations may lead to moieties that are associated with a certain molecular property, training \method{} in an end-to-end fashion circumvents costly first principle calculations. By construction, \method{} allows to identify multiple moieties of the same type (e.g. corresponding to the same functional group) in individual molecules. 
This design choice also makes \method{} transferable w.r.t. molecule size and allows to automatically determine a reasonable number of moieties and environment types without relying on expert knowledge.

Representing molecules as a composition of chemical moieties paves the way for various applications, some of which have been demonstrated in this work: Human-readable and interpretable fingerprints can be directly extracted from the environment types identified by \method{} and used to estimate molecular properties. In addition, they can be employed for selecting representative molecules from quantum mechanical databases to reduce the number of \textit{ab initio} reference data necessary for training accurate models. Further,  we have proposed a CG-MD simulation pipeline based on \method{}, which includes all necessary steps from the identification of CG representations, the machine learning of a CG force field, up to the MD simulation of the CG molecule. The pipeline is fully automatic and does not require expert knowledge. As another example, we have presented the dynamic clustering of chemical reactions, demonstrating that the environment types identified by \method{} capture conformational information that can be used to define reaction coordinates.

A promising avenue for future work is the application of \method{} in the domain of generative models. Jin~et.~al. have shown that generating molecules in a hierarchical fashion can be advantageous~\cite{jin2020hierarchical, jin_junction_2018}. \method{} could help to identify promising motifs for molecule generation and hence facilitate the discovery of large molecules. 
Furthermore, \method{} could be utilized to analyze other interesting reactions. In summary, \method{} extends the applicability of MPNNs to a wide range of chemical problems otherwise relying on expert knowledge. In addition, we expect applications of \method{} to allow new insights into large-scale chemical phenomena, where MPNNs acting on individual atoms are prohibitively computationally expensive to evaluate.

\section*{Acknowledgments}
MG works at the BASLEARN – TU Berlin/BASF Joint Lab for Machine Learning, co-financed by TU Berlin and BASF SE.
KTS, KRM and OTU acknowledge support by the Federal Ministry of Education and Research (BMBF) for the Berlin Institute for the Foundations of Learning and Data (BIFOLD) (01IS18037A).
KRM acknowledges financial support under the Grants 01IS14013A-E, 01GQ1115 and 01GQ0850; Deutsche Forschungsgemeinschaft (DFG) under Grant Math+, EXC 2046/1, Project ID 390685689 and KRM was partly supported by the Institute of Information \& Communications Technology Planning \& Evaluation (IITP) grants funded by the Korea Government(No. 2019-0-00079,  Artificial Intelligence Graduate School Program, Korea University).
OTU acknowledges funding from the Swiss National Science
Foundation (Grant No. P2BSP2\_188147).

\bibliographystyle{ieeetr}
\bibliography{references}

\end{document}


\title{
Automatic Identification of Chemical Moieties\\[2mm]\Large \textsc{(Supplementary Information)}
}

\author{Jonas~Lederer, Michael~Gastegger, Kristof~T.~Schütt\\ Michael~Kampffmeyer, Klaus-Robert M\"uller, Oliver~T.~Unke}

\maketitle
\allowdisplaybreaks



\section{Determining Covalent Bonds from 3D Molecule Structure}
Our approach utilizes RDKit~\cite{ebejer_conformer_nodate} to determine the covalent bonds of each molecule from its 3D structure. This is achieved by constructing a RDKit molecule object from the atomic positions and atomic numbers. Assuming a non-charged molecule, this information is sufficient for RDKit to derive the structural formula of the molecule. Subsequently, the covalent bonds are obtained by applying the module rdkit.Chem.rdmolops.GetAdjacencyMatrix.

\section{Breadth-First Search}\label{section:bfs}
In this section, we describe how to obtain the hard similarity matrix $\mathbf{C}_\mathrm{h}$ (introduced in Section~2.3 of the main text) by utilizing a breadth-first search. Latter is performed on the graph represented by $\mathbf{C}_\mathrm{h}^0$ (defined in eq.~(5) of the main text). In $\mathbf{C}_\mathrm{h}^0$, each atom is set to be similar to its 1st-order neighbors provided that they belong to the same environment type. The 1st-order neighbors are defined by $\mathbf{A}_\mathrm{cov}$ and represent atom-pairs sharing the same covalent bond. The procedure is described by Algorithm~\ref{alg:flip}. 

\begin{algorithm}
\renewcommand{\algorithmicrequire}{\textbf{Input:}}
\renewcommand{\algorithmicensure}{\textbf{Output:}}
\caption{}
    \label{alg:flip}
    \begin{algorithmic}
    \Require $\mathbf{C}_\mathrm{h}^0$
    \vskip 1mm
    \State $\mathbf{C}_\mathrm{h} \gets \mathbf{0}$
    \vskip 1mm
    \State $\mathbf{C}_\mathrm{h}^{\prime} \gets \mathbf{C}_\mathrm{h}^0$ 
    \vskip 1mm
    \While {$\mathbf{C}_\mathrm{h}$ != $\mathbf{C}_\mathrm{h}^{\prime}$}
        \vskip 1mm
        \State $\mathbf{C}_\mathrm{h} \gets \mathbf{C}_\mathrm{h}^{\prime}$
        \vskip 1mm
        \State $\mathbf{C}_\mathrm{h}^\prime \gets \mathbf{C}_\mathrm{h}^0\mathbf{C}_\mathrm{h}$ \Comment{matrix multiplication between $\mathbf{C}_\mathrm{h}^0$ and $\mathbf{C}_\mathrm{h}$}
        \vskip 1mm
        \State $\mathbf{C}_\mathrm{h}^\prime \gets \Theta\left(\mathbf{C}_\mathrm{h}^\prime - \mathbf{1}\right)$ \Comment{set all entries $c$ of $\mathbf{C}_\mathrm{h}^\prime$ to 1, in case $c \geq 1$ and 0 otherwise}
    \EndWhile
    \State \Return $\mathbf{C}_\mathrm{h}$
    \end{algorithmic}
\end{algorithm}

$\Theta$ is the Heaviside step function and $\mathbf{1}$ is a matrix of same dimension as $\mathbf{C}_\mathrm{h}$, where all entries are equal to 1. 
With each iteration the order of included neighbors increases until all atoms of one environment type, which are connected by a set of covalent bonds, are assigned to the same moiety. Hence, after convergence, the resultant matrix $\mathbf{C}_\mathrm{h}$ represents the hard moiety similarity matrix.


\section{Saturation experiment}

\begin{figure}[h]
    \centering
    \includegraphics[width=0.65\textwidth]{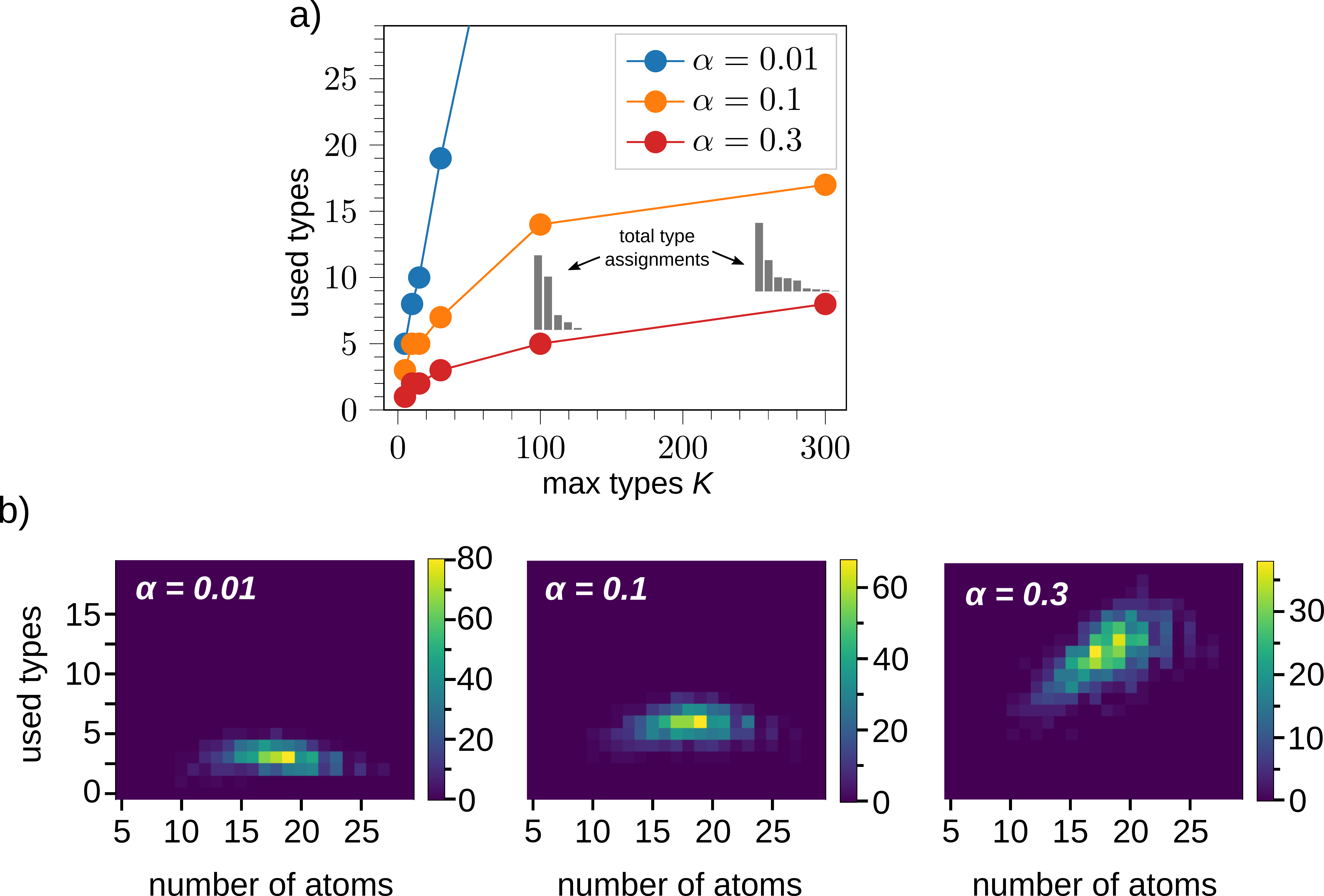}
    \caption{Saturation analysis of the used environment types w.r.t. to the hyperparameters $K$ and $\alpha$. a) The used environment types over the maximum number of types $K$ is depicted for three different values of the entropy trade-off factor $\alpha$. For two runs, additional bar plots indicate the total number of atoms assigned to the respective environment types (each bar represents a single environment type, with its height indicating the number of atoms assigned to that type). b) The number of used environment types depending on the number of atoms in the respective molecules is depicted for three different values of $\alpha$. For all three runs $K=100$.}
    \label{fig:log_entropy}
\end{figure}

The maximum number of environment types $K$ and the entropy trade-off factor $\alpha$ both control the number of formed environment types. To facilitate fine-tuning the model, it is desirable to decouple those parameters such that the number of formed environment types only depends on $\alpha$ for any $K$ above a certain threshold. Figure~\ref{fig:log_entropy}a shows that this is the case for sufficiently large $\alpha$. We see a saturation of environment types with increasing $K$. For small values of $\alpha$, however, the number of formed types is directly proportional to $K$. The reason for this is the decreasing signal to noise ratio with increasing size of the assignment matrix $\mathbf{S}$. For $\alpha=0.1$ and above we consider the number of used types to be independent of $K$ in good approximation. For the runs at $(\alpha=0.3, K=100)$ and $(\alpha=0.3, K=300)$ the respective total environment type assignments are depicted. Each bar represents the amount of atoms assigned to a particular environment type. It can be seen that the last four environment types only exhibit very few atom assignments and the effective number of used types is comparable in both settings.

Figure~\ref{fig:log_entropy}b shows the relation between used environment types and molecule size for a set of 1000 molecules at $K=100$. For $\alpha=0.01$ almost each atom is assigned to its own environment type. With increasing $\alpha$, the number of used types for each molecule becomes less dependent on the molecule size. For $\alpha=0.3$ the number of used types per molecule is almost independent of its size.



\section{Training of SchNet and \method{}}

\subsection{Hyperparameters}
The model-specific hyper-parameters of \method{} are the following:
\begin{itemize}
    \item The cut-off radius $r_\mathrm{min}$ of the min-cut adjacency matrix $\Tilde{\mathbf{A}}$ used in the loss term $\mathcal{L}_\mathrm{cut}$
    \item The cut-off radius $r_\mathrm{bead}$, which determines the distance dependency in similarity matrix $\mathbf{C}$
    \item The entropy trade-off factor $\alpha$.
    \item The maximum number of environment types, which is determined by the cluster dimension $K$ of the type assignment matrix.
\end{itemize}

Choosing $r_\mathrm{min}$ slightly above the covalent bond distance of organic atoms, results in the best representation of the molecular graph. Hence, $r_\mathrm{min}$ does not require any tuning. The cut-off radius $r_\mathrm{bead}$ is associated with the expected size of moieties.

\subsection{Training Set Up}

Both, pretrained \method{} model and end-to-end \method{} model, are trained on the QM9 dataset using the same hyper-parameters $r_\mathrm{bead}=1.8$, $r_\mathrm{min}=3.5$, $\alpha=0.16$, $K=100$. Training set and validation set comprise 11,000 and 1,000 datapoints, respectively, while the remaining points are used for testing. In the case of end-to-end \method{}, a warmup phase (of the cut loss $\mathcal{L}_\mathrm{cut}$ and entropy loss $\mathcal{L}_\mathrm{ent}$ over 50 and 65 epochs, respectively) is added to increase training stability. We utilize the Adam optimizer. The learnable weights, mentioned in Eq.~(1), are initialized at ($\mathbf{W}_1$) uniform and ($\mathbf{W}_2$) orthogonal. For the pretrained \method{} model, the feature representation $\mathbf{X}$ is provided by a pretrained SchNet model. Latter has been trained on the internal energy for 110,000 randomly selected molecules in the QM9 dataset. On a test set of 13,885 molecules, the internal energy is predicted well with a mean absolute error (MAE) of 0.014\,eV.

\subsection{Comparison between Pretrained and End-to-End MoINN}
\begin{figure}[h]
    \centering
    \includegraphics[width=0.49\textwidth]{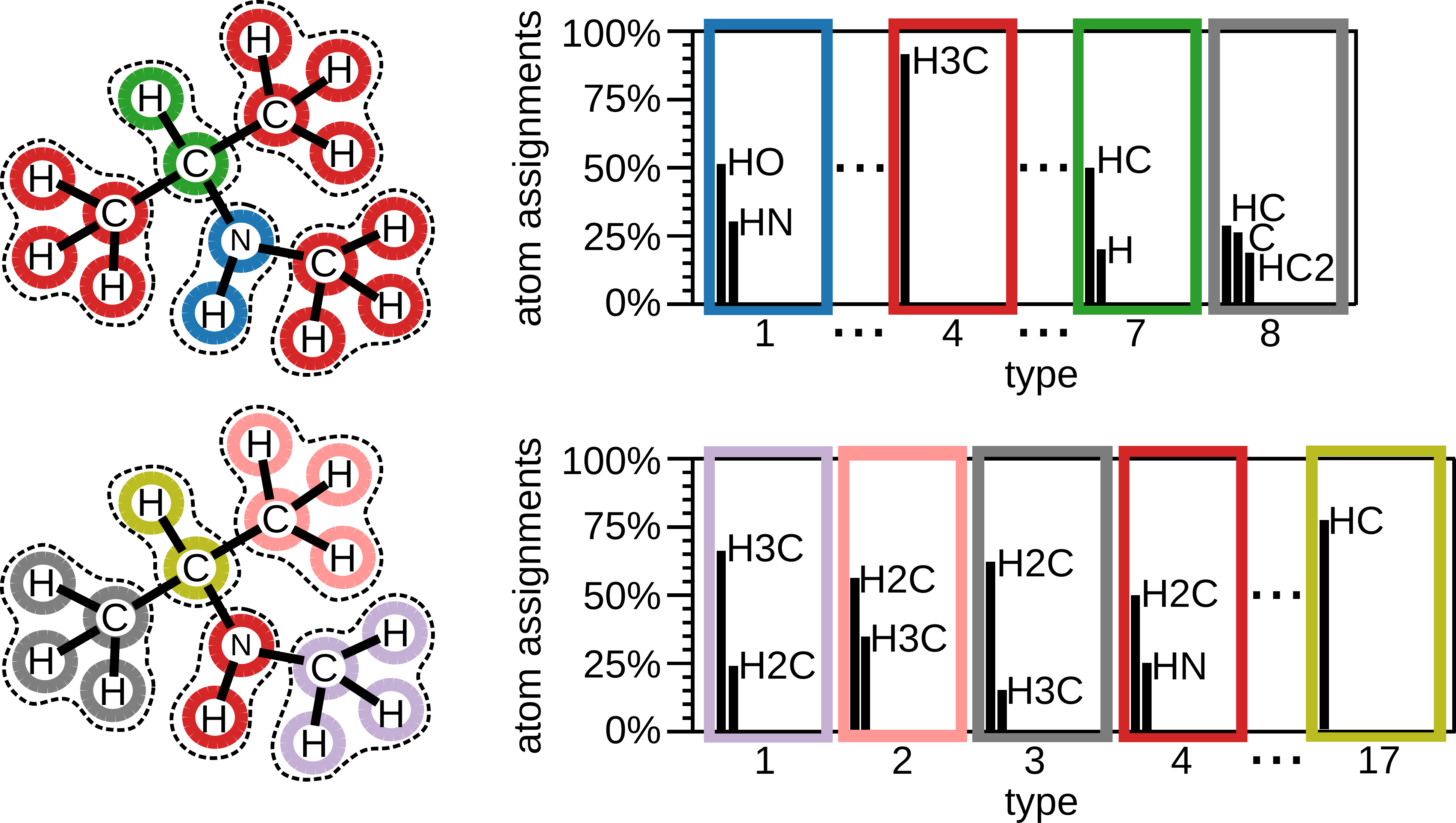}
    \caption{Comparison between (\textbf{top}) pretrained and (\textbf{bottom}) end-to-end MoINN model. For each model the type environments and moieties are depicted (left) for an exemplary molecule and (right) in from of a statistical evaluation based on the test set.}
    \label{fig:moinn_comp}
\end{figure}

Figure~\ref{fig:moinn_comp} depicts the provided results (type environments and moieties) of both, the pretrained model and end-to-end, model. It can be seen that the identified moieties for an exemplary molecule are equivalent for pretrained and end-to-end model. The statistical evaluation shows that for the entire test set, pretrained and end-to-end model identify similar moieties. However, while in the pretrained case, all methyl groups correspond to the same type environment, in the end-to-end case methyl groups in different molecules may be assigned to a different type. This is substantiated by the statistical evaluation, which shows that end-to-end training results in significantly more environment types used, many containing similar moieties. This suggests that training \method{} in an end-to-end fashion results in a more fine-grained division into environment types: Slight variations in the local environment can be captured during the representation learning, while for the pretrained case $\mathbf{X}$ is fixed. 

The advantage of a more fine-grained environment type division is that individual types capture more detailed information about the atomic environments. Hence, the end-to-end model is more suitable for tasks such as data reduction or providing type-based feature representations. However, for applications such as coarse-graining, using a pretrained \method{} allows to represent a greater variety of similar structural motifs with the same environment type. Furthermore, if the aim is extracting chemical insight from the dataset, it may be more natural to use a pretrained \method{} to identify chemical moieties. Intuitively, similar moieties should be associated with similar properties, regardless of their precise (fine-grained) atomic environment, which is better captured with pretrained representations.

\subsection{Training with Varying Training Set Size}

\begin{figure}[h]
    \centering
    \includegraphics[width=0.49\textwidth]{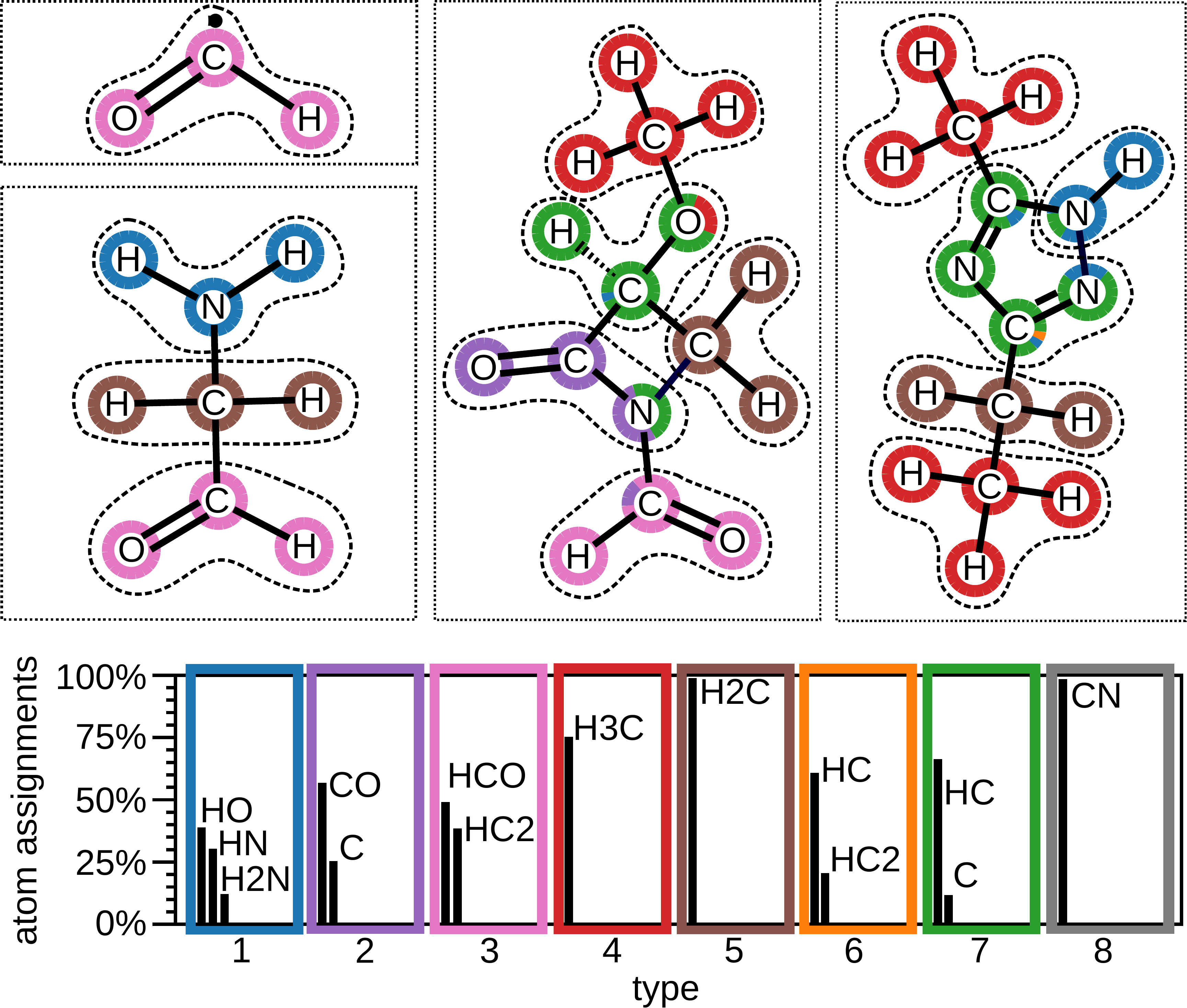}
    \caption{\revision{Common moieties of the QM9 dataset provided by the pretrained MoINN model trained on 110k data points. Equivalent to Fig.~2, the top shows four exemplary molecules along with type assignments (colored circles) and moieties (enclosed by dashed lines). The bottom shows the distribution of environment types and corresponding most common moieties for the test set (1000 molecules), black bars indicate the relative amount of atoms assigned to the respective moieties. For each environment type, over $70\%$ of its atom assignments correspond to at most three different moieties.}}
    \label{fig:qm9_clusters_110k}
\end{figure}

\revision{The MoINN models have been trained on a relatively small set of 11,000 data points, to allow for exhaustive testing on the remaining samples. For completeness, however, we show that a larger training set of 110,000 samples results in very similar results as depicted in Fig.~\ref{fig:qm9_clusters_110k}.} \revision{The environment types of the four exemplary molecules show very strong alignment with those in Fig.~2. Also the statistical distribution of environment types is comparable. Some types as, e.g., the type environment number two (purple) or environment number four (red) are almost identical between the model trained on the large dataset and the model trained on the smaller training set.} 

\section{Detailed Description of Show Cases and Additional Experiments}

\revision{This section provides more detailed descriptions of the show cases in Section~3 as well as some additional experiments to corroborate our findings.}

\subsection{Identification of Chemical Moieties}

\begin{figure}[h]
    \centering
    \includegraphics[width=0.5\textwidth]{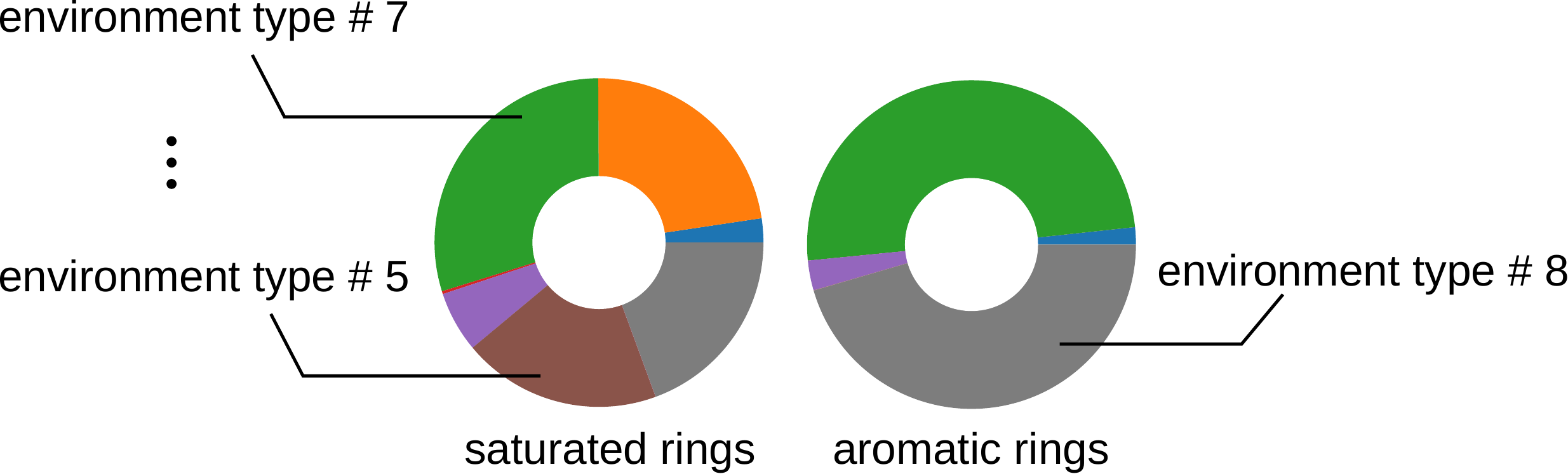}
    \caption{\revision{Average ratio of environment types in saturated and aromatic rings. Each color represents a corresponding environment type.}}
    \label{fig:ring_env_types_statistics}
\end{figure}

\begin{figure}[h]
    \centering
    \includegraphics[width=0.65\textwidth]{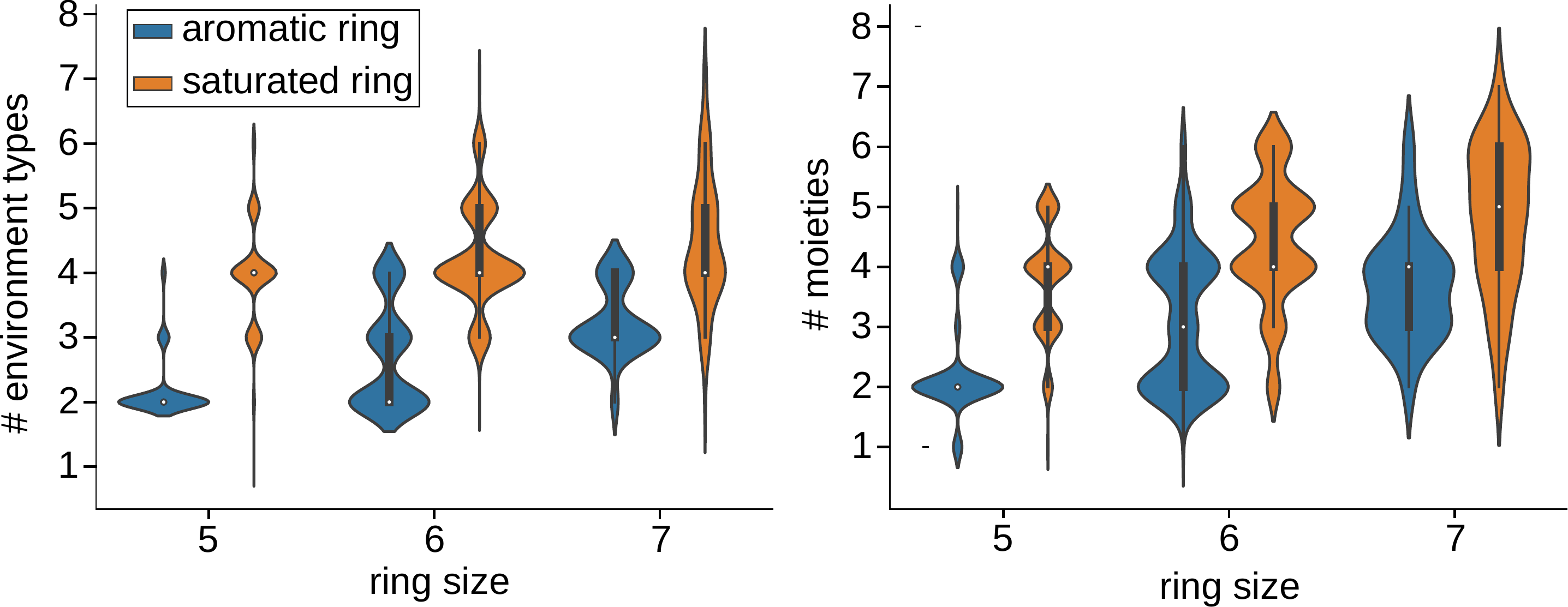}
    \caption{\revision{Comparison between aromatic rings and saturated rings w.r.t. the number of comprised (left) environment types and number of (right) individual moieties. The number of moieties has been determined by the breadth-first search described in Section~\ref{section:bfs}.}}
    \label{fig:ring_statistics}
\end{figure}

\revision{Besides identifying the most common moieties in datasets, MoINN also allows to extract information about more complex substructures such as, e.g., different molecular ring systems. Here we compare the environment types in saturated rings and aromatic rings. Figure~\ref{fig:ring_env_types_statistics} shows the average ratio of environment types in ring systems containing between five and seven heavy atoms. The ratio has been computed for the entire test set of 121,885 samples. It can be clearly seen that atoms in aromatic rings are mostly assigned to two environment types, while saturated rings exhibit several environment types. Figure~\ref{fig:ring_statistics} depicts the respective number of environment types and beads in each ring. It can be seen that aromatic rings tend to exhibit fewer environment types and individual moieties than saturated rings.}

\begin{figure}[h]
    \centering
    \includegraphics[width=0.6\textwidth]{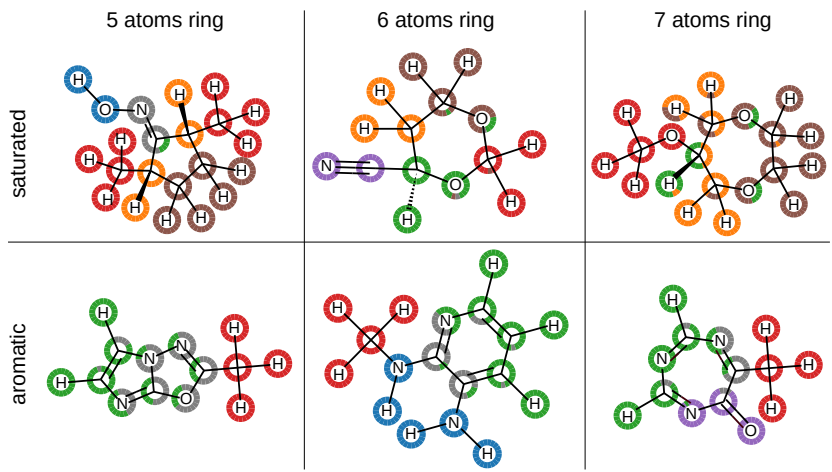}
    \caption{\revision{Six exemplary molecules and their environment type assignments containing saturated and aromatic rings. Rings of three different sizes are compared.}}
    \label{fig:ring_env_types_examples}
\end{figure}

\revision{For a qualitative comparison between saturated and aromatic rings we show some exemplary molecules in Fig.~\ref{fig:ring_env_types_examples}. Equivalent to the quantitative study, the depicted aromatic and saturated rings contain between five and seven heavy atoms. The shown examples corroborate our findings above. Hence we can conclude that MoINN distinguishes between saturated and unsaturated rings. While saturated rings are predominantly divided into several small moieties, aromatic rings are often represented as an individual entity.}

\begin{figure}[h]
    \centering
    \includegraphics[width=0.75\textwidth]{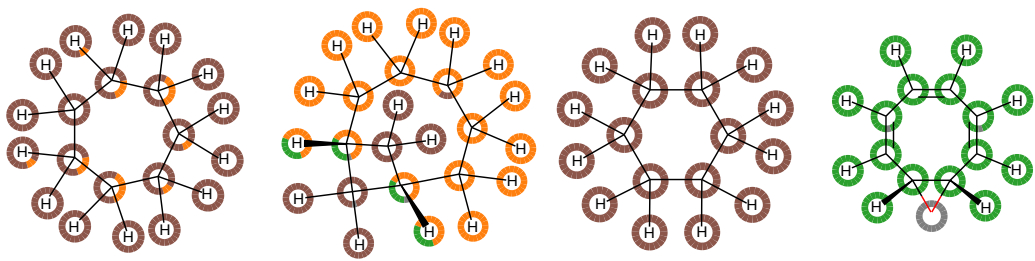}
    \caption{\revision{Largest substructures in QM9 identified by breadth first search based on the environment types of MoINN.}}
    \label{fig:largest_moieties}
\end{figure}

\begin{figure}[h]
    \centering
    \includegraphics[width=0.45\textwidth]{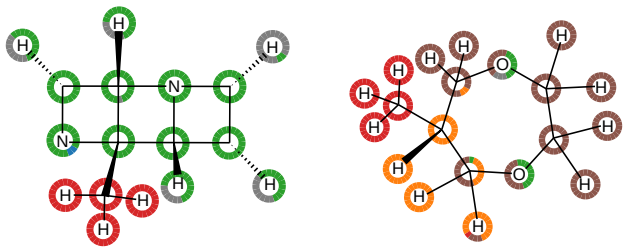}
    \caption{\revision{Large moieties that occur alongside different moieties in the molecule.}}
    \label{fig:largest_moieties_plus_extra_moieties}
\end{figure}

\revision{The most common moieties identified by MoINN (compare Fig.~2) mostly represent small molecular substructures. However, MoINN in combination with the breadth first search (described in Section~\ref{section:bfs}) also identifies larger substructures. The four largest substructures identified in the QM9 dataset are depicted in Fig.~\ref{fig:largest_moieties}. Since the type assignments of MoINN strongly depend on the atomic environments, the largest structures are composed of atoms with very similar atomic environment. Hence, those moeities often comprise the entire molecule. Note that for the task of identifying common moities in the dataset it may be preferable to split up those large substructures into the most common moieties which in this case would be methylene and methine groups (compare Fig.~2). However, also less trivial large substructures are identified. Two of those are shown in Fig.~\ref{fig:largest_moieties_plus_extra_moieties}.} 

\revision{Another approach to verify that the type assignments are chemically meaningful is described in Section~3.1. To this end, we construct molecular fingerprints based on the environment types and compare them to conventional molecular fingerprints by training several linear models on the respective fingerprints. The architecture of those linear models is defined as
$$
\mathbf{y} = \mathbf{XW}
$$
with the trainable weight matrix $\mathbf{W} \in\mathbb{R}^{F\times 1}$, the feature matrix $\mathbf{X} \in\mathbb{R}^{N\times F}$, and the output $\mathbf{y} \in\mathbb{R}^{N}$. $N$ denotes the number of samples and $F$ denotes the feature size. All models are trained in an identical fashion, with a learning rate of $\alpha=10^{-4}$, batch size 100 and the Adam~\cite{DBLP:journals/corr/KingmaB14} optimizer to minimize the loss term which is simply represented by the mean squared deviation between model prediction and the property of the respective sample. The dataset is divided into a training set comprising 110k samples, a validation set of 10k samples and a test set of roughly 22k samples. During the training procedure we use a learning rate scheduler ($\alpha_\mathrm{decay} = 0.8$, $\alpha_\mathrm{patience}= 25$) and early stopping to ensure picking the best performing models for the experiment. The training of all models has converged after at most $500$ epochs. 

The use of a supervised MoINN model for this task does not yield the same level of performance ($\mathrm{MAE}_\mu=\SI{0.11}{Debye}$, $\mathrm{MAE}_H=\SI{54.65}{eV}$, $\mathrm{MAE}_F=\SI{70.79}{eV}$, $\mathrm{MAE}_{U0}=\SI{67.89}{eV}$). In the unsupervised case, incorporating adaptable representations during the training of MoINN, along with the resulting flexibility of the fingerprints, yields better performance.
}

\subsection{Sampling of Representative Molecules}

\begin{figure}[h]
    \centering
    \includegraphics[width=0.4\textwidth]{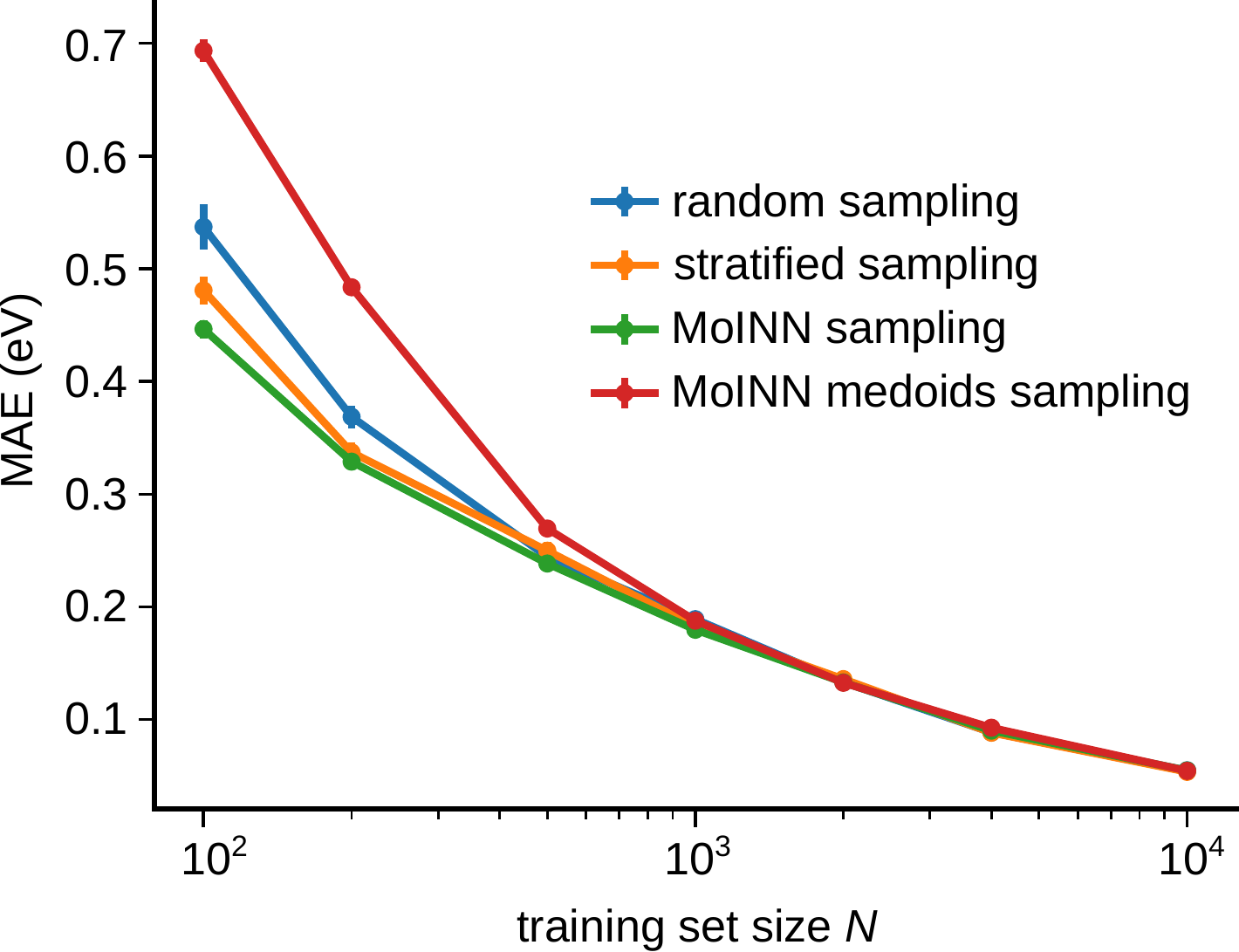}
    \caption{\revision{Mean absolute error (MAE) of energy predictions for SchNet models trained on randomly sampled training sets (blue), training sets obtained by stratified sampling (orange) and training sets selected with MoINN in a self reconstruction manner (green) and using the k-medoids approach (red). Each data point is averaged over five independent training runs and standard errors are indicated by error bars.}}
\label{fig:data_reduction_verbose}
\end{figure}

\revision{

Section~3.2. shows how MoINN can be utilized to extract representative samples from a dataset facilitating the selection of structures for expensive reference calculations. This is achieved by extracting fingerprints from the type environments and minimizing a self reconstruction problem. This way we obtain a small basis set of structures/molecules that represents the dataset. Here we describe the details of this experiment and we extend to experiment to larger data subsets to show that for increasing training set size all approaches converge to similar performance.

The training and validation sets are drawn from a subset of 10,000 samples respectively using the respective methods (random, stratified, MoINN). Table~\ref{tab:data_red} shows the different training and validation set sizes for all runs. For each training set size and validation set size we repeat the procedure five times and train corresponding SchNet models (with the hyperparameters $r_\mathrm{cutoff}=\SI{10}{\angstrom}$, $\mathrm{batch size}=100$, $\mathrm{feature size}=128$, $\mathrm{\#gaussians}=50$, $\mathrm{\#interactions}=3$) for 500 epochs. For the stratified sampling, the dataset is divided into bins, each bin containing molecules of equal size (same number of atoms). Subsequently, the subsets are obtained by uniformly drawing samples from the bins. For MoINN, we compare two approaches, namely, first, the one described in Section~3.2., where we solve a self reconstruction problem and, second, an approach based on medoids~\cite{park2009simple, kaufman2009finding}. The latter finds clusters of MoINN fingerprints and selects $k$ medoids (cluster centers) as representative basis set. The results are shown in Fig.~\ref{fig:data_reduction_verbose}.

\begin{table}[h]
\centering
\small
  \caption{\ Training set and validation set sizes.}
  \label{tbl:example1}
  \begin{tabular*}{0.48\textwidth}{@{\extracolsep{\fill}}lllllll}
    \hline
     & & & \multicolumn{2}{l}{set size}\\
    \hline
    training & $100$ & $200$ & $500$ & $1000$ & $4000$ & $10000$\\
    validation & $100$ & $200$ & $500$ & $1000$ & $1000$ & $1000$\\    
    \hline
  \end{tabular*}
\label{tab:data_red}
\end{table}

}

\revision{
It is evident that up to a training set size of 1000, MoINN sampling provides an advantage for prediction accuracy. As the training set size increases, all methods exhibit similar prediction accuracy. However, utilizing a medoid sampling based on MoINN fingerprints yields worse performance than random sampling, making it an unsuitable candidate for data reduction procedures. The self-reconstruction approach identifies fingerprints that enable the reconstruction of remaining fingerprints through linear combination, whereas the medoids approach identifies fingerprints that are the most dissimilar from each other. This validates our assumption that discovering a basis of MoINN fingerprints is superior to identifying a set of fingerprints that merely represent the variance of the dataset.
}
\subsection{Coarse-Graining}

\begin{figure}
    \centering
    \includegraphics[width=0.5\textwidth]{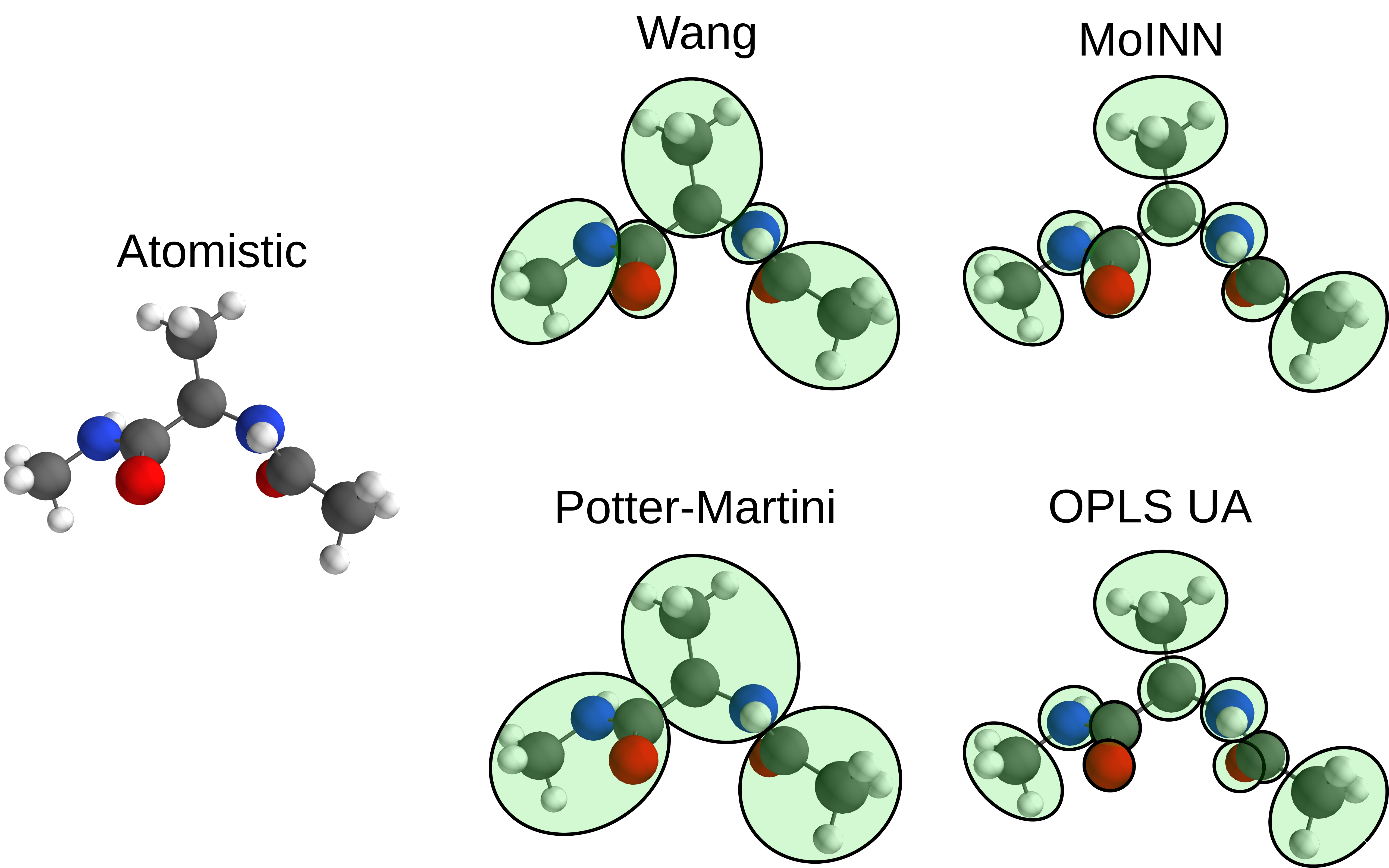}
    \caption{\revision{Comparison between different CG representations provided by Wang~et.~al.~\cite{wang2019machine}, the automated approach for the Martini force field (here referred to as Potter-Martini), our Method MoINN and OPLS UA~\cite{doi:10.1021/ja00214a001, doi:10.1021/ja9621760}.}}
\label{fig:comp_beads}
\end{figure}

\revision{
In Section~3.3 it is shown that MoINN can be utilized to derive CG representations. Here we compare the latter to other CG representations ranging from manually designed representations to automated frameworks. This is depicted in Figure~\ref{fig:comp_beads} where we compare the CG representation provided by MoINN with a CG representation proposed by Wang~et.~al.~\cite{wang2019machine}, the OPLS UA representation introduced by Jorgensen~et.~al.~\cite{doi:10.1021/ja00214a001, doi:10.1021/ja9621760} and the automated coarse graining for the Martini force field~\cite{marrink2007martini} designed by Potter~et.~al.~\cite{potter2021automated} (here referred to as Potter-Martini).

It can be seen that Potter-Martini provides the most coarse representation, followed by Wang, where the molecule is represented by the five backbone atoms. The MoINN representation is a mixture between the latter and the OPLS UA representation. A clear advantage of the CG representation of MoINN is that it provides bead types. This may be particularly useful for beads that exhibit identical compositions but different local environments. We will show this on the example of decaalanine later in this Section.
}

\revision{
To run CG-MD simulations we train several SchNet models. All considered SchNet models are trained for 300 epochs with a batch size of $100$ and a learning rate $\alpha=10^{-5}$. Learning rate scheduler ($\alpha_\mathrm{decay} = 0.8$, $\alpha_\mathrm{patience}= 25$) and early stopping is used to avoid overfitting. The dataset is split into 900k training samples and 100k validation samples. 
For the atomistic SchNet models, we choose a cutoff of $\SI{10.0}{\angstrom}$, feature size $F=128$, $6$ interaction blocks, and a basis expansion of $50$ gaussians. For the coarse-grained model the cutoff is set to $\SI{5.0}{\angstrom}$, we choose feature size $F=128$, $6$ interaction blocks, and a distance expansion of $10$ gaussians. 
}

The force-matching loss function, utilized for training SchNet on the coarse-grained force field, is given by
\begin{equation}
    \mathcal{L} = 
        \rho \left\Vert \hat{U} - U \right\Vert^2 + 
        \frac{1-\rho}{n}\sum_{i=0}^n
            \left\Vert 
                \mathbf{C}_\mathrm{h} \hat{\mathbf{F}}_i + \frac{\partial U}{\partial \mathbf{R}_i^\mathrm{CG}}
            \right\Vert^2~.
    \label{eq:force_matching_loss}
\end{equation}
The trade-off factor is set to $\rho=0.1$. Analogously to the training of an atomistic SchNet model, the environment types defined by \method{} are used to obtain atom type embeddings in the CG SchNet model, i.e., the CG beads are treated as pseudo-atoms. Including the energy error in the loss function is necessary for an ML model that predicts an accurate potential of mean force (PMF). Even though the PMF differs from the potential energy of the atomistic system by definition, taking the energy loss into account with a sufficiently small trade-off factor allows for fitting the forces accurately, while ensuring a reasonable energy difference between the PMF minima. 

\begin{figure}
    \centering
    \includegraphics[width=0.45\textwidth]{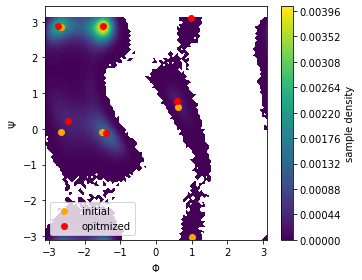}
    \caption{\revision{Density of states of alanine dipeptide projected onto its torsion angles $\phi$ and $\psi$ with indicated free energy minima. The orange (initial) dots correspond to those sates that are associated with the largest sample densities and the red (optimized) dots indicate the states corresponding to the PMF minima.}}
\label{fig:boltzmann_sampling}
\end{figure}

\revision{

For the subsequent CG-MD simulation, we utilize the MD framework provided by SchNetPack~\cite{schutt_schnetpack:_2018}. The latter provides all necessary tools such as integrator, thermostat and logging methods. Our CG-MD simulation comprises 300 trajectories that have been initialized according to the Boltzmann distribution at the six minima of the potential energy surface. The six energy minima are determined based on the density of states in the training dataset. In detail, we select those six conformations that are closest to the maxima of the sample density and perform structure relaxations using the CG SchNet model, respectively. Figure~\ref{fig:boltzmann_sampling} depicts the density projected to the torsion angles $\psi$ and $\phi$ of alanine-dipetide and indicates the sates which represent the minimum energies of the PMF. The Boltzmann distribution
$$
p_i \propto e^{-\epsilon_i/k_\mathrm{B}T}
$$
describes the probability of the physical system to be in a certain state $i$ with the corresponding energy $\epsilon_i$. Here we sample at room temperature $T=\SI{300}{\kelvin}$ and $k_\mathrm{B}$ denotes the Boltzmann constant. Starting from $300$ initial states, we run MD simulations in the NVT setting for $\SI{8}{\nano\second}$ with an integration step of $\SI{2}{\femto\second}$. The thermal bath provided by the Langevin thermostat is updated each $100$ steps. 
}

\begin{figure*}[tb]
    \centering
    \includegraphics[width=\textwidth]{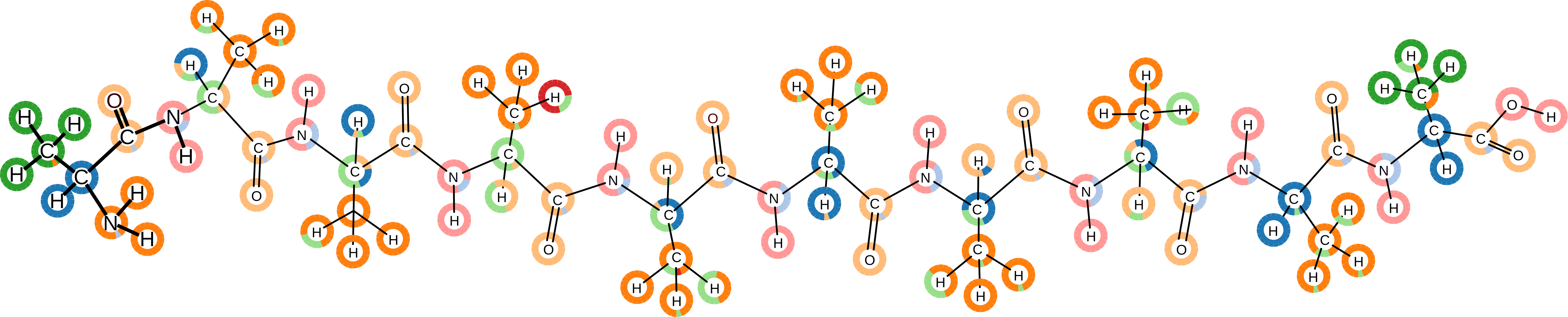}
    \caption{\revision{Environment types for decaalanine provided by \method{}.}}
\label{fig:decaalanine}
\end{figure*}

\revision{
In our work, we show on the example of alanine-dipeptide that MoINN can be employed to find CG representations of molecules. However, MoINN can be applied to molecules of any size due to its transferability with respect to the number of atoms. Figure~\ref{fig:decaalanine} depicts the environment type assignments for decaalanine provided by an end-to-end MoINN model which has solely been trained on small molecules from QM9 ($ \# \textrm{heavy atoms} \le 9$). For large molecules, the larger number of different environment types associated with end-to-end \method{} models (see Section~S4\dag), results in more meaningful moiety representations.

Similar to the case of alanine-dipeptide, the environment types can be utilized to define a coarse-grained representation of the molecule. However, molecules with $ \# \textrm{heavy atoms} \gg 9$ are likely to exhibit some atomic environments that strongly deviate from those in the QM9 dataset. This explains some undesired behaviour such as, e.g., assigning $\mathrm{NH}_2$ and $\mathrm{CH}_3$ to the same type or single carbon atoms being assigned to individual beads. Hence, finding CG-beads using only the automatic breadth first search algorithm is not recommended. Nevertheless, the identified environment types resonate with chemical intuition and tremendously facilitate selecting CG-beads. In this case, when defining the CG representation, we mainly rely on the automated breadth first search process with a few exceptions: (i)~Beads are only assigned the same type if they comprise the same composition of atom types. (ii)~Individual hydrogen atoms are assigned to their nearest heavy atoms. The respective bead then gets the type of the heavy atom. (iii)~Only atoms, which are connected by covalent bonds, can be pooled to the same bead. As mentioned above, CG representations based on MoINN have the advantage that bead types are provided. In the case of decaalanine we can see that the terminal methyl groups are assigned to a different type then the methyl groups at the backbone of the molecule. This may facilitate learning an appropriate force field for this molecule representation.
}

\subsection{Dynamic Clustering and Reaction Coordinate Analysis}
\begin{figure}
    \centering
    \includegraphics[width=0.45\textwidth]{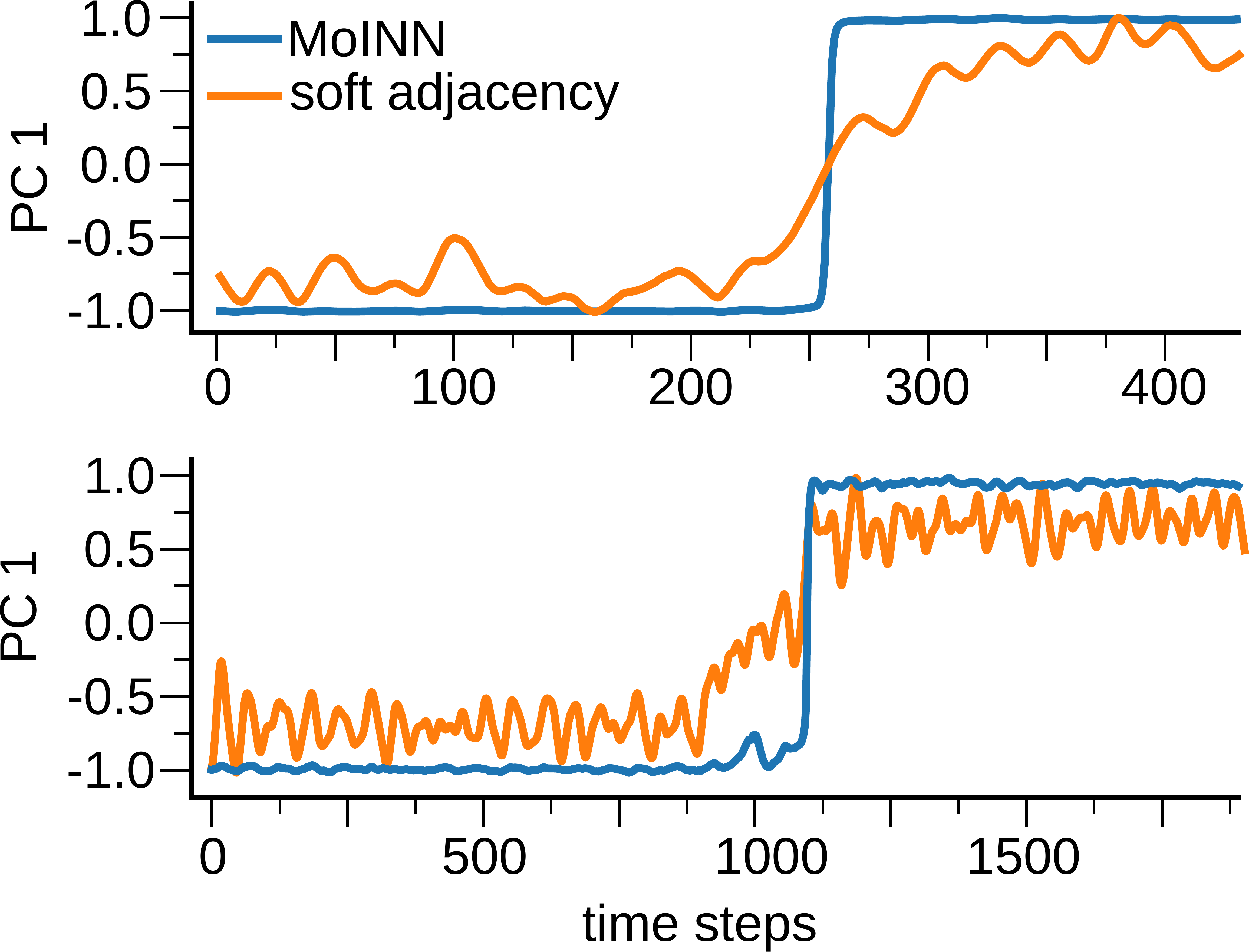}
    \caption{\revision{Reaction coordinates for (\textbf{top}) proton transfer in malondialdehyde and (\textbf{bottom}) Claisen rearrangement based on MoINN and the soft adjacency matrix, respectively.}}
\label{fig:reaction_coordinate_plus_adj}
\end{figure}

\revision{As described in Section~3.4, we can extract reaction coordinates from the type assignments provided by MoINN. Since the variation of the structure during the reaction is also covered in the pairwise distances between atoms, also dimensionality reduction of the adjacency matrix should provide a reasonable reaction coordinate. The reaction coordinate based on MoINN is derived as described in Section~3.4. Similarly, for the soft adjacency, we define the reaction coordinate by the first principle component of the flattened adjacency matrix. Equivalent to the min-cut adjacency matrix, used in MoINN, we choose a Cosine cutoff function with the cutoff radius $r_\mathrm{cut}=\SI{1.8}{\angstrom}$ to calculate the adjacency of atom pairs. 

In Figure~\ref{fig:reaction_coordinate_plus_adj}, we compare reaction coordinates for malondialdehyde and the Claisen rearrangement, on the one hand based on MoINN and on the other hand relying on the adjacency matrix. We observe that, as expected, both reaction coordinates allow to distinguish between reactant and product state. However, MoINN provides a sharper representation of the state transition, while the reaction coordinate based on the soft adjacency matrix appears noisy. 
}

\section{Limitations of Common Graph-Pooling Methods}

In this section, we describe the graph-pooling methods MinCUT Pooling and DiffPool, and emphasize their limitations w.r.t. the clustering of molecules. Both approaches utilize a soft assignment matrix for coarsening graphs. They both introduce auxiliary loss terms to ensure a finite number of localized clusters. For the case of molecules, we will show that adapting their proposed loss terms allows for more reasonable clustering. Both approaches allow for hierarchical graph-pooling. For our desired applications, however, considering a single pooling step is sufficient, which is why we do not expand on the hierarchical features of MinCUT and DiffPool.

\subsection{Comprehensive Discussion of MinCUT Pooling}

The concept of minCUT pooling was first stated by Bianchi~et.~al~\cite{bianchi_mincut_2019}. It describes the acquisition of an assignment matrix ${\mathbf{S}\in\mathbb{R}^{N\times K}}$ which is used to link $N$ graph nodes to their respective $K$ clusters. $\mathbf{S}$ is also often referred to as affinity matrix and is given by 
\begin{equation}
    \mathbf{S} = \mathrm{softmax}\left(
        \mathrm{ReLU}\left(\mathbf{XW}_1\right)\mathbf{W}_2
    \right)~.
    \label{eq:mincut_ass_layer}
\end{equation}
$\mathbf{W}_1\in\mathbb{R}^{F\times H}$ and $\mathbf{W}_2\in\mathbb{R}^{H\times K}$ are trainable weights matrices, with the hidden dimension $H$, and $\mathbf{X}$ is the feature representation. Eventually, applying the \textit{softmax} function ensures that all cluster assignments of each row  obey $\sum_j^K s_{ij}=1$ with ${s_{ij}>0}$. Thus, $\mathbf{s}_i$ represents the cluster assignment probability distribution of the \textit{i}th node.

In addition to the task-specific supervised loss, an unsupervised loss is minimized. Latter is given by
\begin{equation}
    \mathcal{L} = 
        -\frac{Tr\left(\mathbf{S}^T\tilde{\mathbf{A}}\mathbf{S}\right)}{Tr\left(\mathbf{S}^T\tilde{\mathbf{D}}\mathbf{S}\right)} + 
        \left\Vert 
            \frac{\mathbf{S}^T\mathbf{S}}{\left\Vert\mathbf{S}^T\mathbf{S}\right\Vert_F} - \frac{\mathbf{I}_K}{\sqrt{K}}
        \right\Vert_F~.
    \label{eq:mincut_clustering_loss}
\end{equation}
$\tilde{\mathbf{A}}=\mathbf{D}^{-1/2}\mathbf{AD}^{-1/2}\in\mathbb{R}^{N\times N}$ is the symmetrically normalized adjacency matrix of the molecular graph, where $\mathbf{D}\in\mathbb{R}^{N\times N}$ denotes the degree matrix, which is a diagonal matrix with elements $d_{ii} = \sum_j^N a_{ij}$. There, $\{a_{ij}\}$ are the entries of the adjacency matrix. Consequently, $\tilde{\mathbf{D}}$ is the degree matrix of $\tilde{\mathbf{A}}$. $\mathbf{I}_K\in\mathbb{R}^{N\times N}$ is the identity matrix, and $\Vert \cdot \Vert_F$ is the Frobenius norm. The first term in \eqref{eq:mincut_clustering_loss} is denoted as the cut loss term $\mathcal{L}_\text{c}$ and favours clusters of adjacent nodes. To avoid converging to the trivial solution of $\mathcal{L}_\text{c}$ which corresponds to assigning all nodes to the same single cluster, a second term is used in \eqref{eq:mincut_clustering_loss}. It ensures that the assignment vectors are close to orthogonal and hence it is referred to as the orthogonality loss $\mathcal{L}_\text{o}$. This rewards assignments associated with clusters of balanced size. For more details refer to ~\cite{bianchi_mincut_2019}.

\subsection{Comprehensive Discussion of DiffPool}
DiffPool was proposed by Ying~et.~al.~\cite{ying_hierarchical_2018}. The assignment matrix is given by
\begin{equation}
    \mathbf{S} = \mathrm{softmax}\left( \mathrm{GNN}_\mathrm{pool}\left( \mathbf{A}, \mathbf{X} \right) \right)~,
    \label{eq:diffpool_ass}
\end{equation}
where $\mathrm{GNN}_\mathrm{pool}$ is a graph neural network. 
Similar to the MinCUT loss, DiffPool uses an auxiliary unsupervised loss, which reads
\begin{equation}
    \mathcal{L} = \left\Vert \mathbf{A} - \mathbf{SS}^T\right\Vert_F - \frac{1}{N}\sum_{nk} S_{nk} \ln{(S_{nk})}~.
    \label{eq:diffpool_loss}
\end{equation}
The first term is called {\it{Auxiliary Link Prediction Objective}}, and it favours localized clusters of nodes, analogously to MinCUT's cut loss. The second term, the {\it{Entropy Regularization}}, is minimized, when the cluster assignments represent one-hot vectors. This avoids the trivial solution of assigning all nodes to a single cluster. Hence this loss term is similar to the orthogonality loss in MinCUT.

\subsection{The Issue of Symmetries in Molecules}

The MinCut and DiffPool approaches are designed to avoid assigning distant nodes to the same cluster. However, \eqref{eq:mincut_ass_layer} and \eqref{eq:diffpool_ass} link the atomic representations to their assignments, such that nodes with a similar environment exhibit similar cluster assignments. Hence, the approaches assume that distant nodes exhibit different feature representations. This is not necessarily the case for molecules, which may be highly symmetric, leading to similar feature representations of distant nodes (atoms). 

\bibliographystyle{ieeetr}
\bibliography{references}